\documentclass[a4paper,11pt]{article}
\usepackage{jheppub}
\usepackage{amsmath}
\usepackage{slashed}
\usepackage{amssymb}
\usepackage{color}
\usepackage{graphicx}
\usepackage{hyperref}
\usepackage{xspace,slashed}
\usepackage[utf8]{inputenc}
\usepackage{subcaption}
\usepackage{multirow}
\usepackage{algorithm}
\usepackage{algorithmic}
\usepackage{stackengine}
\usepackage{enumitem}
\usepackage{etoolbox}
\usepackage{wasysym}

\newtoggle{draft}

\togglefalse{draft}

\iftoggle{draft} {
  \usepackage{changes}
}{%
  \usepackage[final]{changes}
}

\definechangesauthor[name=pn, color=blue]{pn}
\definechangesauthor[name=km, color=magenta]{km}

\usepackage{booktabs}

\makeatletter
\g@addto@macro\bfseries{\boldmath}
\makeatother

\newcommand{\bp}{{\bf p}}
\newcommand{\bk}{{\bf k}}

\newcommand{\figref}[1]{Fig.~\ref{#1}}

\newcommand{\be}{\begin{equation}}
\newcommand{\ee}{\end{equation}}
\newcommand{\ep}{\varepsilon}
\newcommand{\nn}{\nonumber \\}

\newcommand\scalemath[2]{\scalebox{#1}{\mbox{\ensuremath{\displaystyle #2}}}}

\usepackage{soul}
\allowdisplaybreaks

\newcommand{\TTPaff}{Institute for Theoretical Particle Physics,
  KIT, 76128 Karlsruhe, Germany}

\preprint{
  \begin {flushright}
    TTP23-017, P3H-23-032
  \end{flushright}
}

\title{
  Non-factorizable virtual corrections
  to Higgs boson production in weak boson fusion beyond the eikonal approximation
}
\author[]{Ming-Ming Long,}
\author[]{Kirill Melnikov,}
\author[]{J\'er\'emie Quarroz}
\affiliation[]{\TTPaff}

\emailAdd{ming-ming.long@kit.edu}
\emailAdd{kirill.melnikov@kit.edu}
\emailAdd{jeremie.quarroz@kit.edu}

\abstract{
  Non-factorizable virtual corrections to Higgs boson production in weak boson
  fusion at next-to-next-to-leading order in QCD
  were  estimated in the eikonal approximation \cite{Liu:2019tuy}. This approximation
  corresponds
  to the expansion of relevant amplitudes around the forward limit.  In this paper we compute the
  leading  power correction to the eikonal limit and show that it is proportional
  to  \emph{first power} of the
  Higgs boson  transverse momentum or the Higgs boson mass over  partonic center-of-mass
  energy.  Moreover, this correction can be significantly enhanced by the rapidity of the
  Higgs boson. For realistic weak boson fusion cuts, the next-to-eikonal correction reduces
  the estimate of non-factorizable contributions  to fiducial  cross section
  by  ${\cal O}(30)$ percent. 
    }

\begin{document}

\maketitle 
\section{Introduction}

At the Large Hadron Collider (LHC),  Higgs bosons are
frequently produced in weak boson fusion (WBF).
This process has a recognizable signature, characterized by two energetic low-$p_\perp$
jets in the opposite  hemispheres.
Higgs boson production in WBF has been measured  by CMS~\cite{CMS:2015ebl,CMS:2018uag} and
ATLAS~\cite{ATLAS:2018jvf,ATLAS:2019nkf} collaborations. 
The measured WBF cross section agrees with the Standard Model prediction to within $20\%$.
Further improvements in the experimental exploration of  Higgs boson production in weak boson
fusion are expected during the Run~III and the high-luminosity phase of the LHC. 

Theoretical understanding  of Higgs boson production in WBF is very advanced. It is based on the knowledge  
of next-to-leading order (NLO)~\cite{Figy:2003nv,Berger:2004pca} and
next-to-next-to-leading order (NNLO)~\cite{Bolzoni:2010xr,Bolzoni:2011cu,Cacciari:2015jma,Cruz-Martinez:2018rod} QCD corections, as well
as mixed QCD-EW~\cite{Ciccolini:2007ec} corrections to this process. 
N$^3$LO QCD corrections have also been computed~\cite{Dreyer:2016oyx}; they
change the leading-order cross section by just about one permille. 

It is to be noted, however, that all these studies were performed in the so-called  factorization approximation 
where contributions due to  gluon exchanges between two incoming  fermion lines are neglected. 
These effects, that we will refer to  as \emph{non-factorizable corrections},
are color-suppressed  and, for this reason, are expected to be smaller than the factorizable ones \cite{Bolzoni:2010xr,Bolzoni:2011cu}.
However, virtual non-factorizable corrections, which start contributing to the WBF cross section  at NNLO QCD,
exhibit a peculiar enhancement by two powers of $\pi$. This enhancement was first
observed when the two-loop non-factorizable
amplitude was computed  in the leading eikonal approximation~~\cite{Liu:2019tuy}.
To better understand these two-loop effects and to
establish the validity of the eikonal approximation for
phenomenological analyses of Higgs boson production in weak boson fusion,
it is essential to go beyond the leading term in the eikonal expansion.

Since  the calculation of exact non-factorizable contributions, which requires 
the two-loop five-point amplitude with five independent kinematic variables and two masses,
is currently not possible, 
it is reasonable to explore the possibility to extend the eikonal expansion beyond the forward limit. 
In this paper we make the first step in that direction and compute the leading 
power correction to the eikonal limit of  non-factorizable five-point WBF amplitude.

The remainder of this paper  is organized as follows. In the next  section, we describe  kinematics
of weak boson fusion  and explain how we use it to  set up an  expansion around the eikonal limit.
In Sections~\ref{sec:one_loop_calc} and \ref{sec:two_loop_calc},
we derive integral representations  for  one- and two-loop amplitudes which contribute to non-factorizable
corrections to WBF; these representations retain  the next-to-eikonal accuracy. In Section~\ref{sec:pole_cancel},
we explain how the infra-red finite,  two-loop non-factorizable correction can be derived from these integral representations.
In Section~\ref{sec:num} we  analyze the numerical impact of the computed next-to-eikonal corrections and show that
they change the current estimate of the non-factorizable contribution
to the WBF cross section by about ${\cal O}(30)$ percent.
We conclude in Section~\ref{sec:conclusion}.  Discussion of the analytic computation
of one- and two-loop non-factorizable amplitudes is relegated to appendix. The analytic results for the amplitudes
can be found in an ancillary file provided with this submission. 

\section{Kinematics of Higgs production in weak boson fusion}
\label{sec:kinematics}

We begin with the  discussion of the  kinematics of Higgs production in the  WBF   process
\be
q(p_1) + q(p_2) \to q(p_3) + q(p_4) + H(p_H)\,.
\label{eq:def_kin}
\ee
We perform the Sudakov decomposition of the four-momenta of the outgoing quarks and write 
\be
\begin{split}
  & p_3 = \alpha_3 p_1 + \beta_3 p_2 + p_{3,\perp}\,,
  \\
  & p_4 = \alpha_4 p_1 + \beta_4 p_2 + p_{4,\perp}\,.
\end{split} 
\ee
Employing  the on-shell conditions  $p^2_3=0$, $p^2_4=0$, we find\footnote{Throughout the paper,
  the bold-faced
  notation is used for two-dimensional Euclidian vectors.}
\be
\begin{split} 
\beta_3 = \frac{{\bf p}_{3,\perp}^2}{s \alpha_3}\,,\;\;\;\;
  \alpha_4 = \frac{{\bf p}_{4,\perp}^2}{s \beta_4}\,,
\end{split} 
\ee
where $s = 2 p_1 \cdot p_2$ is the partonic center-of-mass energy squared. 
The WBF events are selected by requiring  that two tagging jets
with a relatively small transverse momentum are present in opposite hemispheres;
this ensures that $\alpha_3 \sim \beta_4 \sim 1$ and that
${\bf p}_{3,\perp}^2 \sim {\bf p}_{4,\perp}^2 \ll s$.

  We define two auxiliary vectors $q_1$ and $q_2$  which describe momentum  transfers from the 
  quark lines  to the Higgs boson. They read
  \be
  \begin{split}
    & q_1 = p_1 - p_3 = \delta_3  p_1 - \beta_3 p_2 - p_{3,\perp}\,,
    \\
    & q_2 = p_2 - p_4 = -\alpha_4 p_1 + \delta_4 p_2 - p_{4,\perp}\,,
    \end{split} 
  \ee
where $\delta_3 = 1-\alpha_3$ and $\delta_4 = 1-\beta_4$. 
  It follows from the momentum conservation condition that 
  \be
p_H = q_1 + q_2\,. 
  \ee
  Upon squaring the two sides of this equation and some rearrangements, we find
  \be
  \delta_3 \delta_4 s = m_H^2 + \frac{{\bf p}_{3,\perp}^2}{\alpha_3}
  + \frac{{\bf p}_{4,\perp}^2}{\beta_4} +2 {\bf p}_{3,\perp} \cdot {\bf p}_{4,\perp} - \frac{{\bf p}_{3,\perp}^2 {\bf p}_{4,\perp}^2}{\alpha_3 \beta_4 s}\,.
  \label{eq2.6}
  \ee

  We can use Eq.~(\ref{eq2.6}) to fully  specify the relevant aspects of WBF kinematics around the forward limit.
  Indeed, given the proximity of the Higgs boson mass  and electroweak boson masses, and the fact that
  the important contribution to WBF cross section comes from kinematical configurations where the transverse momenta of tagging jets 
  are comparable to  $m_H$ and $m_{W,Z}$, the above equation implies
  \be
  \delta_3 \delta_4  \sim \frac{m_V^2}{s} \sim \frac{m_H^2}{s} \sim
  \frac{\bp_{3,\perp}^2}{s} \sim \frac{\bp_{4,\perp}^2}{s} \sim \lambda \ll 1\,.
  \ee
  Note that we introduced a  parameter  $\lambda$
  to indicate  the smallness of various ratios in the above equation. 
  We consider central production of Higgs bosons  so that   neither forward nor backward direction is preferred. Then 
   $\delta_3 \sim \delta_4$ and 
  \be
 \delta_3 \sim \delta_4 \sim \sqrt{\lambda} \gg \lambda\,.
 \ee

 We note that, with the required accuracy,  the two parameters $\delta_{3,4}$ can be written
 as follows
 \be
\delta_{3,4} = \sqrt{ \frac{ \bp_{H,\perp}^2 + m_H^2}{s}}   e^{\pm y_H}\,,
 \ee
 where $ \bp_{H,\perp} $ is the transverse momentum and $y_H$ is the rapidity of the Higgs boson in the partonic center-of-mass frame. 
 We will use the above relations between kinematic parameters to construct the expansion of one- and
 two-loop non-factorizable WBF amplitudes in the following sections.

\section{One-loop non-factorizable contributions to WBF}
\label{sec:one_loop_calc}

We consider the
one-loop non-factorizable QCD corrections to Higgs boson production in  WBF. To avoid confusion, we note that
they do not
contribute to the WBF cross section at NLO  since their interference with the leading order
amplitude vanishes because of  color conservation. Nevertheless, since
the one-loop amplitude is needed for the construction of the NNLO QCD corrections, we need to discuss it.

To write the non-factorizable amplitude in a convenient way, we 
assume that the coupling of the vector boson $V$  to the Higgs boson is
given by $i g_{VVH} \; g_{\mu \nu} $ and that  the coupling of the massive vector boson to quarks
is vector-like, $-ig_W \gamma^\mu$.  Since we work with massless quarks, their helicities
are conserved and we can reconstruct non-factorizable
contributions for  $V = Z$ and $V = W$ from the results that are  reported  below. 

\begin{figure}[t]
\begin{center}
  \begin{subfigure}{0.48\textwidth}
    \centering
    \includegraphics[height=3.5cm]{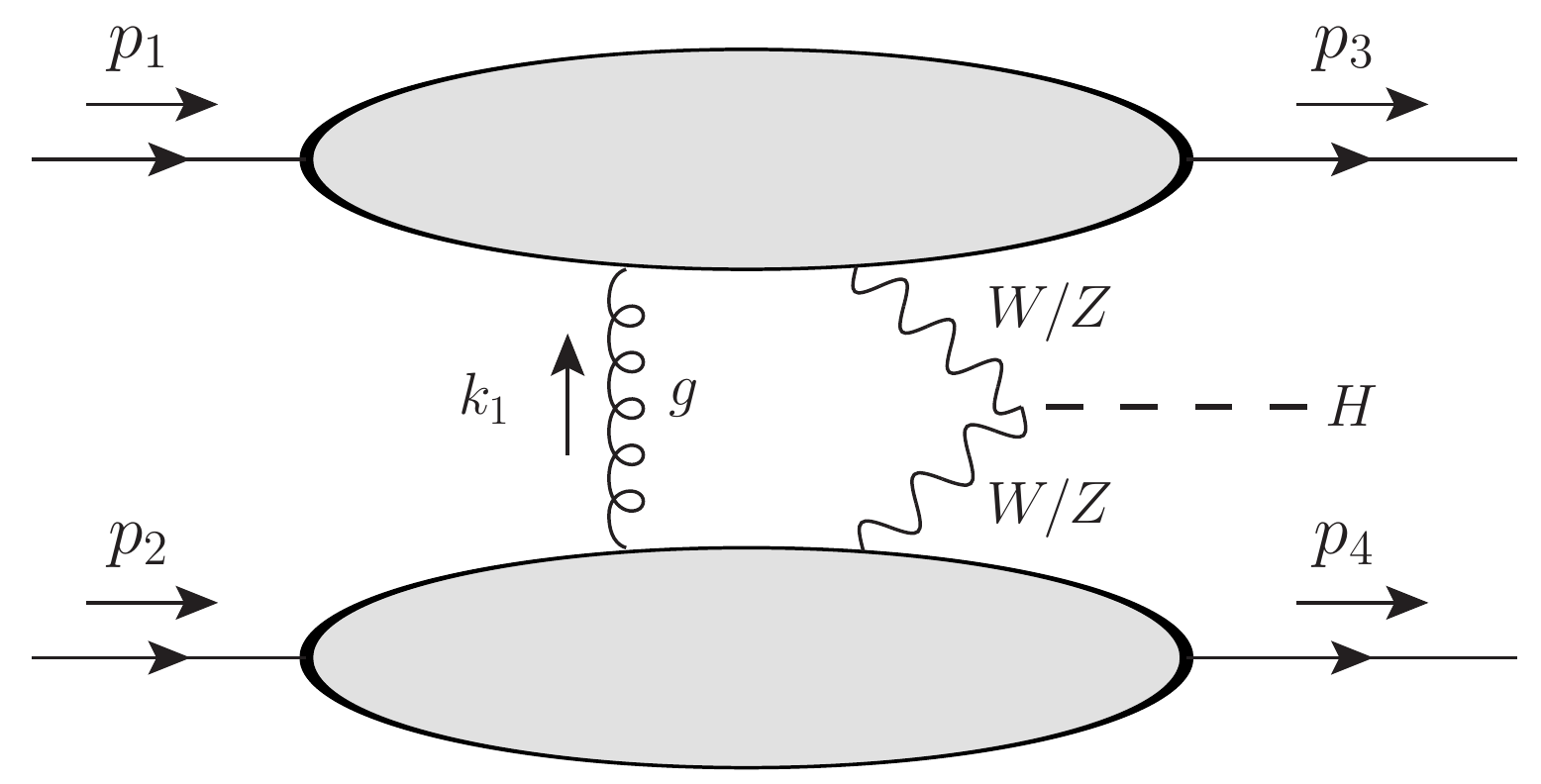}
    \label{fig:wbf_1l}
  \end{subfigure}
  ~
  \begin{subfigure}{0.48\textwidth}
    \centering
    \includegraphics[height=2.8cm]{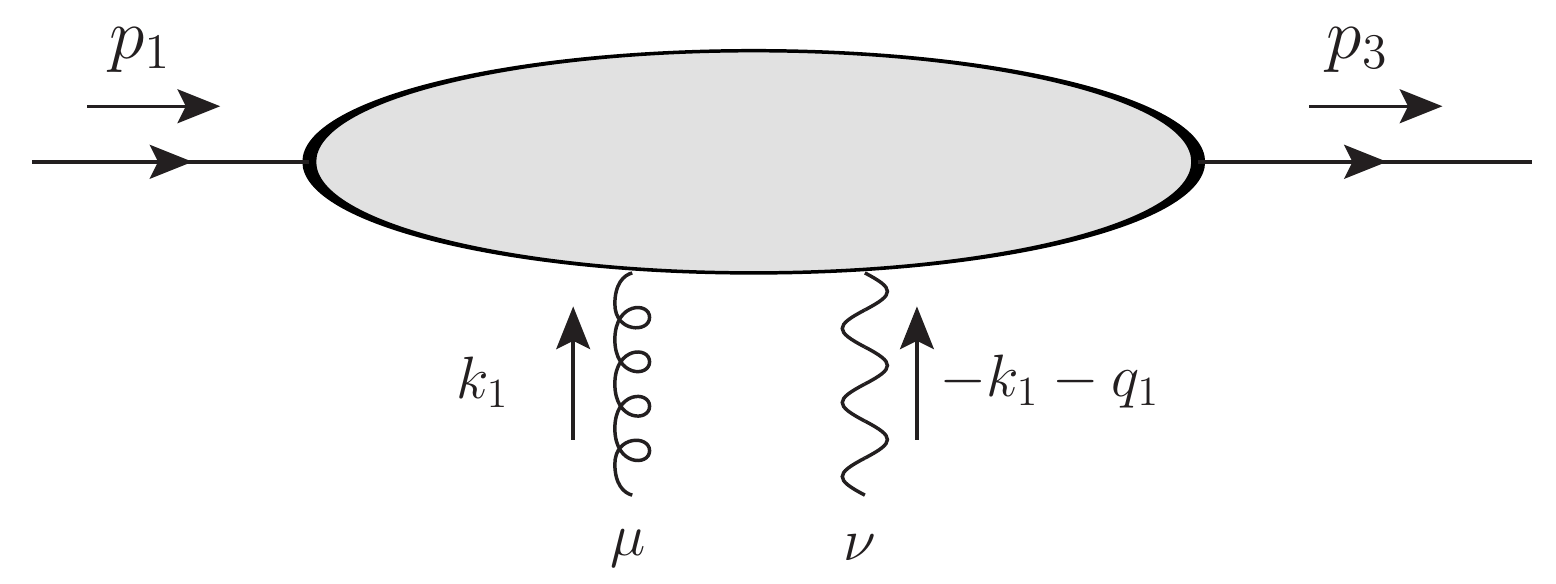}
    \label{fig:current_NLO}
  \end{subfigure}
  \caption{The one-loop amplitude, shown on the left, can be constructed by contracting the currents
    for the upper and lower fermion lines. The current for the upper fermion
    line $J^{\mu \nu}(k_1,-k_1-q_1)$  is shown  on the right. 
 }
\label{fig0}
\end{center}
\end{figure}

We write the one-loop non-factorizable amplitude as follows
\be
   {\cal M}_{1} = g_s^2 g_W^2 \; g_{VVH} \; T^{a}_{i_3 i_1} T^a_{i_4 i_2} \; {\cal A}_{1}\,,
   \label{eq:oneloop_start}
   \ee
   where $T^{a}_{ij}$ denote  the generators of the
   $SU(3)$ color group and  ${\cal A}_{1}$ stands for  the color-stripped one-loop amplitude\footnote{Throughout this
     paper, we use dimensional regularization, with the dimensionality of space-time being 
     $d = 4 - 2\ep$.} 
   \be
{\cal A}_1  =  \int \frac{{\rm d}^d k_1}{(2\pi)^d } \frac{1}{d_1 d_3 d_4}
   J_{\mu \nu}(k_1,-k_1-q_1) \; {\tilde J}^{\mu \nu}(-k_1,k_1-q_2)
   \,.
     \label{eq3.1}
\ee
In Eq.~\eqref{eq3.1}, we used the notation 
\be
d_1 = k_1^2+i0, \;\;\; d_3 = (k_1 + q_1)^2 - m_V^2 + i0, \;\;\; d_4 = (k_1-q_2)^2 - m_V^2 + i0
\,,
\ee
to define propagators of virtual bosons. In addition, following the conventions  in \figref{fig0}, we introduced
two quark currents
\be
\begin{split} 
  & J^{\mu \nu}(k_1,-k_1-q_1) = \langle 3|
  \left [ \frac{\gamma^\nu ( \hat p_1 + \hat k_1) \gamma^\mu }{\rho_1(k_1)} + \frac{\gamma^\mu (\hat p_3 - \hat k_1) \gamma^\nu }{\rho_3(-k_1)}
    \right ] |1 ]\,,
    \\
  & {\tilde J}^{\mu \nu}(-k_1,k_1-q_2) = \langle 4|
  \left [ \frac{\gamma^\nu (\hat p_2 + \hat k_1) \gamma^\mu }{\rho_2(k_1)} + \frac{\gamma^\nu (\hat p_4 - \hat k_1) \gamma_\mu }{\rho_4(-k_1)}
    \right ] |2 ]\,,
  \end{split}
\label{eq3.2}
\ee
where we assumed that the incoming fermions are left-handed. 
In writing Eq.~(\ref{eq3.2}) we employed  the quantities $\rho_i(k)$, $i=1,2,3,4$ to describe  quark propagators;
they read
\be
\rho_i(k) = \frac{1}{(p_i+k)^2 + i0}\,.
\ee

\vspace*{0.5cm}
We would like to construct an expansion of the amplitude in Eq.~(\ref{eq3.1}) in powers of $\lambda$. To understand how
to do tatt, we introduce the Sudakov parametrization of the loop momentum $k_1$ and write
\be
k_1 = \alpha_1 p_1 + \beta_1 p_2 + k_{1,\perp}\,. 
\ee
The integration measure in Eq.~(\ref{eq3.1}) becomes
\be
\frac{{\rm d}^d k_1 }{(2\pi)^d} = \frac{s}{2} \;  \frac{ {\rm d} \alpha_1}{2 \pi} \;  \frac{ {\rm d} \beta_1}{2 \pi}
 \frac{{\rm d}^{d-2} \bk_{1,\perp} }{(2 \pi)^{d-2}}\,.
\ee

\begin{table*}[t]
\centering
\begin{tabular}{|c|c|c|c|}
\hline
Region    & $\alpha_1$ & $\beta_1$  & $\bk_{1,\perp}$   \\
\hline
\hline 
a & $\lambda $ & $\lambda$ & $\sqrt{ \lambda} $ \\
b & $\lambda  $ & $ \sqrt{\lambda} $ & $\sqrt{\lambda} $ \\
c & $ \sqrt{\lambda} $ & $ \sqrt{\lambda} $ & $\sqrt{\lambda} $ \\
d & $ 1  $ & $ \lambda $ & $\sqrt{ \lambda} $ \\
e & $1  $ & $ 1 $  & $ 1 $ \\
\hline
\end{tabular}
\caption{Kinematic regions relevant for one-loop non-factorizable contributions. Symmetric
regions are not shown.}
\label{tab1}
\end{table*}

The various propagators in Eq.~(\ref{eq3.1}) are linear polynomials in $\alpha_1$ and $\beta_1$.
Hence, integration over either one of these two variables can be easily  performed using the residue theorem.
The resulting integrand is a  product of (at most)  quadratic polynomials in the other variable so that
the structure of singularities can be easily analyzed.  Performing this analysis and assuming that
the transverse loop momentum can either be of the same order as 
 the transverse momenta of the outgoing jets or  of the same order as  the center-of-mass energy, 
 we come to the conclusion that
 the following loop-momenta regions,\footnote{See
   Refs.~\cite{Beneke:1997zp,Jantzen:2011nz,Jantzen:2012mw} for the discussion of the strategy of regions
   and its application to computing loop integrals.}    shown in Table~\ref{tab1}, need to be considered.
The first region is the so-called Glauber region; the  second one is   ``Glauber-soft'',  the third
one is soft, the fourth is collinear and the last one is hard.

Using the scaling of the  loop-momentum components as indicated in Table~\ref{tab1}, we estimate  
the contributions of the various
regions to the one-loop amplitude.  We find 
\be
   {\cal M}^{(a)} \sim \lambda^{-2}\,,
   \;\;\;    {\cal M}^{(b)} \sim \lambda^{-2}\,,
   \;\;\; {\cal M}^{(c)} \sim \lambda^{-2}\,, \;\;\; {\cal M}^{(d)} \sim \lambda^{-3/2}\,,
   \;\;\; {\cal M}^{(e)} \sim 1\,.
   \label{eq3.8}
\ee
We note that the leading order WBF amplitude scales as $\lambda^{-2}$ and that,
as  follows from Eq.~(\ref{eq3.8}),  the expansion of the one-loop
amplitude proceeds in powers of $\sqrt{\lambda}$.
To compute  ${\cal O}(\sqrt{\lambda})$  correction
to the virtual amplitude, we need to account for the  contributions of regions $a)$, $b)$ and $c)$ to  first subleading
power and the contribution  of region  $d)$ to leading power   in the expansion in $\lambda$. 

\vspace*{0.5cm}
We begin with the discussion of region $a)$. Using momentum scaling in Table~\ref{tab1}, we simplify the various
propagators that appear in the integrand in Eq.~(\ref{eq3.2}).   To present the result in a compact way, we
introduce the following quantities 
\be
\begin{split}
  & \Delta_1 = -\bk_{1,\perp}^2,\;\;\; \Delta_{3,1} = - (\bk_{1,\perp} - \bp_{3,\perp})^2 - m_V^2,\;\;\; \Delta_{4,1} = - (\bk_{1,\perp} + \bp_{4,\perp})^2 - m_V^2\,,
  \\
  & \Theta_{3,1} = - \left ( \bk_{1,\perp}^2 - 2 \bk_{1,\perp} \cdot \bp_{3,\perp} \right )\,,\;\;\;\;\;\;
  \Theta_{4,1}  = - \left ( \bk_{1,\perp}^2 + 2 \bk_{1,\perp} \cdot \bp_{4,\perp} \right )\,.
\end{split}
  \ee
  In region $a)$, all inverse propagators scale as ${\cal O}(\lambda)$. To compute the first subleading correction
  we need to keep all terms that scale as $\lambda^{3/2}$ and neglect all terms that scale as $\lambda^2$.
  We find 
\be
\begin{split} 
  & d_1  \approx  \Delta_1 + i0,\;\;\;  d_3 \approx  s \delta_3 (\beta_1 - \beta_3) + \Delta_{3,1} + i0,
  \;\;\; d_4 \approx  -s \delta_4 (\alpha_1 + \alpha_4) +  \Delta_{4,1}  + i0\,,\;\;
    \\
  & \rho_1(k_1) \approx  s \beta_1  + \Delta_1 + i0\,,
  \;\;\;\;\;\;\;\;\;\;\;\;\; \rho_2(-k_1) \approx -s \alpha_1 + \Delta_1  + i0\,,
 \\
 & \rho_3(-k_1) \approx  - s \alpha_3 \beta_1 + \Theta_{3,1}  + i0\,,
 \;\;\; \rho_4(k_1) \approx  s \beta_4 \alpha_1 + \Theta_{4,1}  + i0\,.
 \label{eq3.9}
\end{split} 
\ee
If we use the simplified propagators shown in Eq.~(\ref{eq3.9}) to compute the
amplitude ${\cal A}_1$,  we observe that integrations over $\alpha_1$ and $\beta_1$ factorize. We then write 
  \be
     {\cal A}_1^{(a)}
  =
     -\frac{s}{2} \;   \int
     \frac{{\rm d}^{d-2} \bk_{1,\perp} }{(2 \pi)^{d-2}} \; \frac{1}{\Delta_1 \Delta_{3,1} \Delta_{4,1} }  \; \Phi^{\mu \nu}  \; \tilde \Phi_{\mu \nu}
     \,,
     \label{eq3.11}
\ee
where 
\begin{align}
\begin{split}
&\Phi^{\mu \nu} =  \int  \limits_{-\sigma}^{\sigma} \frac{ {\rm d} \beta_1}{2 \pi i}
     \frac{\Delta_{3,1}}{s \delta_3 (\beta_1 - \beta_3) + \Delta_{3,1} +i0 } \\
& \qquad \qquad \qquad \times  \langle 3|
       \left [ \frac{\gamma^\nu ( \hat p_1 + \hat k_{1,\perp} ) \gamma^\mu }{s \beta_1 + \Delta_1  + i0  }
            + \frac{\gamma^\mu (\hat p_3 - \hat k_{1,\perp} ) \gamma^\nu }{- s \alpha_3 \beta_1 + \Theta_{3,1} + i0 }
            \right ] |1 ]\,,
\end{split}
\\
\begin{split}
& \tilde \Phi^{\mu \nu} =  \int \limits_{-\sigma}^{\sigma}  \frac{ {\rm d} \alpha_1}{2 \pi i}
\frac{\Delta_{4,1} }{-s \delta_4 (\alpha_1 + \alpha_4) + \Delta_{4,1} +i0} \\
& \qquad \qquad \qquad \times \langle 4|
       \left [ \frac{\gamma^\nu (p_2 + \hat k_{1,\perp} ) \gamma^\mu }{-s \alpha_1 + \Delta_1 + i0  }
            + \frac{\gamma^\nu (p_4 - \hat k_{1,\perp} ) \gamma^\mu }{ s \beta_4 \alpha_1 + \Theta_{4,1}  + i0 }
            \right ] |2 ]\,.
         \label{eq3.13}
\end{split}
\end{align}
         In Eq.~(\ref{eq3.11})  $\sigma$ is a cut-off parameter that forces $\beta_1$ and $\alpha_1$ to stay in the region $\alpha_1 \sim \beta_1
         \sim \lambda $.  It is convenient to choose $\sigma$ such that
         \be
 \lambda \ll \sigma \ll \sqrt{\lambda}
 \,, 
         \ee
         since this choice  will allow us to use the same cut-off $\sigma$ to study the Glauber-soft region.

         We note that  we  replaced $\hat k_1$ with  $\hat k_{1,\perp}$ in
         the currents when  writing Eq.~(\ref{eq3.11}); this is justified 
         since
         $\alpha_1$ and $\beta_1$ terms in the Sudakov expansion of $k$ provide
         ${\cal O}(\lambda)$ and not ${\cal O}(\sqrt{\lambda})$ corrections in region $a)$.  Hence, if we aim at computing the 
         non-factorizable amplitude with ${\cal O}(\sqrt{\lambda})$ relative accuracy, we can discard them.
         In fact,  to compute the amplitude with  ${\cal O}(\sqrt{\lambda})$ relative accuracy,
         terms with $\hat k_{1,\perp}$ in Eq.~(\ref{eq3.11})
         can be dropped altogether. Indeed, since 
 $k_{1,\perp} \sim \sqrt{\lambda}$,
         if we retain it in one of the terms that appear  either in   $\Phi^{\mu \nu}$ or in $\tilde \Phi^{\mu \nu}$, the other current
         should be computed at leading $\lambda$-power.  However, in this  case 
         \be
         \langle 4 | \gamma^\mu \hat p_{2,4} \gamma^\nu |2 ] \approx 4 p_2^\mu p_2^\nu\,, \;\;\;\;\;\;
           \langle 3 | \gamma^\mu \hat p_{3,1} \gamma^\nu |1 ] \approx 4 p_1^\mu p_1^\nu\,, \;\;\;\;\;
         \ee
         and terms with $ \hat k_{1,\perp}$ lead to the vanishing contributions 
         \be
\hat p_i \; \hat k_{1,\perp} \; \hat p_i = 0,\;\;\; i = 1,2\,, 
         \ee
since  $p_{1,2}^2 = 0$ and $p_{1,2} \cdot k_{1,\perp} = 0$. 

Furthermore, in region $a)$  we can expand the
remnants of weak boson propagators that appear in Eq.~(\ref{eq3.11}).  Keeping   terms that 
provide ${\cal O}(\sqrt{\lambda})$ corrections,  we find 
         \be
\begin{split} 
        &  \frac{\Delta_{3,1}}{s \delta_3 (\beta_1 - \beta_3) + \Delta_{3,1} + i0}
         \approx 1 + \frac{s \delta_3 (\beta_3 - \beta_1) }{\Delta_{3,1}} + {\cal O}(\lambda) \,, \\
         & \frac{\Delta_{4,1}}{-s \delta_4 (\alpha_1 + \alpha_4) + \Delta_{4,1} + i0}
         \approx  1 + \frac{ s \delta_4 (\alpha_4 + \alpha_1)}{\Delta_{4,1}} + {\cal O}(\lambda)\,.
         \label{eq3.15}
\end{split} 
         \ee
         Focusing on  $\Phi^{\mu \nu}$, we simplify the expression for the current, use Eq.~(\ref{eq3.15}) and
         obtain\footnote{We note that we are allowed to discard
           $\bp_{3,\perp}$ from the numerator in the expression for $\Phi^{\mu \nu}$
	   for the same reason that   $k_{1,\perp}$ was  discarded.}
         \be
         \Phi^{\mu \nu} =  \frac{ 2 p_1^\mu \langle 3 | \gamma^\nu |1 ]}{s}\;  \Phi\,,
           \ee
           where
           \be
           \begin{split} 
             \Phi = \int  \limits_{-\sigma}^{\sigma}
             \frac{ {\rm d} \beta_1}{2 \pi i} \left ( 1 + \frac{s \delta_3 (\beta_3 - \beta_1) }{\Delta_{3,1}} \right )
        &   \Bigg  [\frac{1}{ \beta_1 + \frac{\Delta_1}{s}  + i0  } + \frac{1}{-  \beta_1 + \frac{\Theta_{3,1} }{s \alpha_3} + i0 }
            \Bigg  ]\,.
          \end{split} 
         \ee
         To compute $\Phi$, we use 
         \be
         \int  \limits_{-\sigma}^{\sigma}
         \frac{ {\rm d} \beta_1}{2 \pi i} \; \frac{1}{ \pm  \beta_1 - z_a + i 0} = -\frac{1}{2} + {\cal O}(z_a/\sigma)
         \,,
\ee
valid for $z_a \in [-\sigma,\sigma]$.  Furthermore, we need
\be
\int_{-\sigma}^{\sigma}  \frac{ {\rm d} \beta_1}{2 \pi i} \; \beta_1 \; \left ( \frac{1}{ \beta_1 - z_a + i 0}
+ \frac{1}{ -\beta_1 - z_b + i 0}
\right) 
= -\frac{1}{2} \left ( z_a - z_b \right ) +{\cal O}(z_a^2/\sigma,z_b^2/\sigma)
\,.
\ee
Neglecting the $\sigma$-dependent terms that will cancel with the contribution from the Glauber-soft region,
we  obtain 
\be
\Phi = (-1) 
\left [ 1  + \frac{ \delta_3}{2 \Delta_{3,1}} \left ( 2 s\beta_3 + \Delta_1 - \Theta_{3,1} \right )
  \right]\,.
\ee
A similar computation for $\tilde \Phi^{\mu \nu}$ gives 
\be
\tilde \Phi^{\mu \nu} = \frac{ 2 p_2^\mu  \langle 4| \gamma^\nu | 2]}{s} \tilde \Phi\,,
  \ee
  where 
\be
\tilde \Phi =  (-1) 
\left [ 1  + \frac{ \delta_4}{2 \Delta_{4,1}}
   \left ( 2s\beta_4 + \Delta_1 - \Theta_{4,1} \right ) \right ]\,.
\label{eq:one_loop_phi_tilde}
\ee
           Combining these results for $\Phi$ and $\tilde \Phi$
           and neglecting all terms beyond desired ${\cal O}(\sqrt{\lambda})$ corrections,
           we obtain the following contribution to the one-loop amplitude from the Glauber region
           \be
\begin{split} 
           {\cal A}_1^{(a)}
              & = 
              - \langle 3 | \gamma^\mu | 1] \langle 4| \gamma_\mu | 2]
                  \;   \int
		  \frac{{\rm d}^{d-2} \bk_{1,\perp} }{(2 \pi)^{d-2}} \; \frac{1}{\Delta_1 \Delta_{3,1} \Delta_{4,1} }
\\
                 & \times  \left (
1 + \frac{ \delta_3}{2 \Delta_{3,1}} \left ( 2 s\beta_3 + \Delta_1 - \Theta_{3,1} \right )
+  \frac{ \delta_4}{2 \Delta_{4,1}}
   \left ( 2s\beta_4 + \Delta_1 - \Theta_{4,1} \right ) \right )
   \,.
\label{eq3.23}
\end{split}
\ee\\

We then proceed with the discussion of the contribution of region $b)$ with 
the mixed  scaling  $\alpha_1 \sim \lambda$ and $ \beta_1 \sim \sqrt{\lambda}$.
 According to Eq.~(\ref{eq3.8}),
 we require the contribution of this region through  first subleading terms. However, it is easy to see 
 that, in actuality,  the contribution of region  $b)$ starts  at ${\cal O}(\lambda^{-3/2})$ and, therefore,
 should be computed at \emph{leading power} only.

 To understand why this is the case, we first discuss the currents $J^{\mu \nu}$ and ${\tilde J}^{\mu \nu}$ and,
 in particular, the numerators of the contributing terms. Since we work with ${\cal O}(\sqrt{\lambda})$ accuracy,
 in region $b)$ we should replace $k_1$ with $k_1 \to \beta_1 p_2 + k_{1,\perp}$ in \emph{both} currents.
 Suppose we do this replacement in $J^{\mu \nu}$. Since these terms already provide an ${\cal O}(\sqrt{\lambda})$
 correction, the current ${\tilde J}^{\mu \nu}$ should be taken at leading power. Since at leading power
 ${\tilde J}^{\mu \nu} \sim p_2^\mu p_2^\nu$, it is easy to see that all contributions of vector $k_1$ drop
 from the current $J^{\mu \nu}$ once the Lorentz indices are contracted.

 However, if we account for
 $k_1$ in the current ${\tilde J}^{\mu \nu}$,  the situation is different. In this case, since $i)$ $k_1$ is 
 independent of $\alpha_1$, $ii)$ it appears with different signs in the two terms in ${\tilde J}^{\mu \nu}$,
 and $iii)$ ${\tilde J}^{\mu \nu}$ is contracted with  ${J}^{\mu \nu}$ computed at leading power,
 the corresponding contribution vanishes after integration over $\alpha_1$. 

 Having concluded that, similar to the Glauber region, we can drop  $k_1$ from the fermion currents,
 we note that  the current $J^{\mu \nu}(k_1,-q_1-k_1)$ in region $b)$ can be further simplified. 
Indeed, using the fact that $\beta_1 \gg \Delta_{1}/s, \Theta_{3,1}/s$, we expand the current and obtain
 \be
 \begin{split} 
J^{\mu \nu}(k_1,-q_1-k_1) & \approx p_1^\mu p_1^\nu
\left (  \frac{1}{s \beta_1 + \Delta_1 +
  i0} + \frac{ \alpha_3}{-s\alpha_3 \beta_1 + \Theta_{3,1} + i0} \right )
\\
& \approx  -\frac{p_1^\mu p_1^\nu}{s \beta_1^2} \left (\Delta_1 + \Theta_{3,1} \right )
\,.
 \end{split}
 \label{eq3.25}
 \ee
 This equation implies that in region $b)$ the current scales as ${\cal O}(1)$ and not as ${\cal O}(\lambda^{-1/2})$
 as a naive estimate suggests. This suppression  occurs
 because of the cancellation between two terms in brackets  in Eq.~(\ref{eq3.25}).
 This  means that the contribution of the region $b)$ starts at $\lambda^{-3/2}$,  so that
 all  ingredients needed to compute the amplitude in region $b)$, except the current $J^{\mu \nu}(k_1,-q_1-k_1)$, are to be 
 taken at leading power in $\lambda$. 

 Hence, we find
 \be
    {\cal A}_{1}^{(b)} = - \langle 3 | \gamma_\mu | 1] \langle 4 | \gamma^\mu | 2]
        \int 
	\frac{{\rm d}^{d-2} \bk_{1,\perp} }{(2 \pi)^{d-2}} \; \frac{1}{\Delta_1 \Delta_{3,1} \Delta_{4,1} }\;
        \Delta \Phi \; \tilde \Phi, 
        \ee
	where $\tilde \Phi$ is still given by Eq.~(\ref{eq:one_loop_phi_tilde}) and 
        \be
        \Delta \Phi = \left (-\frac{\Delta_1}{s} - \frac{\Theta_{3,1}}{s}  \right )
        \int \limits_{-\infty}^{\infty} 
        \frac{{\rm d} \beta_1}{2 \pi i} \frac{( \theta(\beta_1 - \sigma) +\theta(-\sigma - \beta_1) )\Delta_{3,1}}{ ( s \delta_3 \beta_1 + \Delta_{3,1} + i 0)
          \; \beta_1^2}
          \,.
        \ee
        Calculation of this integral is straightforward. We obtain\footnote{We do not display contributions
          that  scale as $\Delta_{3,1}/\sigma$ since they cancel against the contribution of the Glauber region.}
        \be
\Delta \Phi = \frac{\delta_3}{2 \Delta_{3,1}} \left ( \Delta_{1} + \Theta_{3,1}  \right )\,.
\ee
Performing a similar computation for a symmetric region $\beta \sim \lambda, \; \alpha \sim \sqrt{\lambda}$,
we obtain 
\be
\Delta \tilde \Phi = \frac{\delta_4}{2 \Delta_{4,1}} \left ( \Delta_{1} + \Theta_{4,1}  \right )\,.
\ee
Combining the contributions of regions $a)$ and $b)$, we find
\be
\begin{split} 
{\cal A}_1^{a \& b}
 & = 
              - \langle 3 | \gamma^\mu | 1] \langle 4| \gamma_\mu | 2]
                  \;   \int
		  \frac{{\rm d}^{d-2} \bk_{1,\perp} }{(2 \pi)^{d-2}} \; \frac{1}{\Delta_1 \Delta_{3,1} \Delta_{4,1} }
\\
                 & \times  \left (
1 + \frac{ \delta_3}{ \Delta_{3,1}} \left ( s\beta_3  - \Theta_{3,1} \right )
+  \frac{ \delta_4}{\Delta_{4,1}}
   \left ( s\alpha_4  - \Theta_{4,1} \right ) \right )\,.
\end{split}
\label{eq3.30}
\ee

\vspace*{0.5cm}
We turn our attention to   region
$c)$ which  corresponds to the soft scaling
$\alpha_1 \sim \beta_1 \sim |\bk_{1,\perp}| \sim \sqrt{\lambda}$. According to Eq.~(\ref{eq3.8})
we require the contribution of this region through  first subleading power. However,
a more careful analysis shows that   the contribution
of this region is suppressed stronger than originally expected.
To see this we note that in the soft region, to leading power, the currents
           \emph{vanish}.  For example, the  expression for $J^{\mu \nu}(k_1,-q_1-k_1)$ reads
             \be
           J^{\mu \nu}(k_1,-q_1-k_1) \approx p_1^\mu p_1^\nu
             \left (  \frac{1}{s \beta_1 + i0} + \frac{ \alpha_3}{-s\alpha_3 \beta_1 + i0} \right )
              = p_1^\mu p_1^\nu ( -2 i \pi) \delta(\beta_1) \to 0\,, 
              \label{eq:ignore_pole}
           \ee
           and we have set it to zero because poles of the fermion propagators have already been
           accounted for  when the Glauber
           region was analyzed.   Hence, to obtain a non-vanishing contribution from the soft region, subleading terms
           in \emph{both} currents $J^{\mu \nu}$ and ${\tilde J}^{\mu \nu}$ are needed.
           The subleading contributions to the currents
           scale as ${\cal O}(1)$ and not as $1/\sqrt{\lambda}$ as a naive estimate for the currents'
           scaling  would suggest. This
           implies that at  variance with the original
           estimate ${\cal M}^{(c)} \sim \lambda^{-2}$ in Eq.~(\ref{eq3.8}), the contribution
           of the soft region is suppressed by an additional power of $\lambda$. For this reason, the soft
           region is not needed  for computing the two-loop non-factorizable
           amplitude with ${\cal O}(\sqrt{\lambda})$ accuracy. 

           The contribution of the collinear region can be analyzed in the same way.
           Since, in this case, the amplitude scales
           as ${\cal M}^{d} \sim \lambda^{-3/2}$, both currents need to be taken at leading power.
           We find 
\be
\begin{split} 
  & J^{\mu \nu}(k_1,k_2) = \langle 3|
  \left [ \frac{\gamma^\nu ( \hat p_1 + \beta_1 \hat p_2 ) \gamma^\mu }{\beta_1 s + i 0}
      + \frac{\gamma^\mu (\hat p_1  - \beta_1 \hat p_2 ) \gamma^\nu }{-\beta_1 s + i 0}
      \right ] |1 ]
    \\
    & =
    \langle 3 |  \gamma^\mu \hat p_2 \gamma^\nu  + \gamma^\nu \hat p_2 \gamma^\mu  |1 ]
      = 2  \langle 3 | p_2^\mu \gamma^\nu + p_2^\nu \gamma^\mu - g^{\mu \nu} \hat p_2  |1 ],
           \\
  & {\tilde J}^{\mu \nu}(k_1,k_2) = \langle 4|
  \left [ (1 + \beta_1)   \frac{\gamma^\nu \hat p_2 \gamma^\mu }{\rho_2(k_1)} + (1-\beta_1) \frac{\gamma^\nu  \hat p_2 \gamma^\mu }{\rho_4(-k_1)}
    \right ] |2 ]\,.
\end{split}
\label{eq3.32a}
\ee
It is clear that the contraction of the two currents in Eq.~(\ref{eq3.32a}) 
vanishes. Hence,
we conclude that collinear regions do not provide the ${\cal O}(\sqrt{\lambda})$ corrections
to the leading term in the eikonal expansion.
  Since, obviously, the hard region is not relevant as well, we conclude that,
with ${\cal O}(\sqrt{\lambda})$ accuracy, 
the one-loop non-factorizable contribution  is given  by the sum of the Glauber and
Glauber-soft contributions  in Eq.~(\ref{eq3.30}).

Having performed this analysis, we note that the final result for the two regions $a)$ and $b)$ can be
obtained by simply computing the functions $\Phi$ and $\tilde \Phi$ from the following unexpanded expressions 
          \be
           \begin{split} 
             & \Phi = \int  
             \frac{ {\rm d} \beta_1}{2 \pi i} \frac{\Delta_{3,1}}{
               s \delta_3 (\beta_1 - \beta_3) + \Delta_{3,1}+i 0}
                \Bigg  [\frac{1}{ \beta_1 + \frac{\Delta_1}{s}  + i0  }
               + \frac{1}{-  \beta_1 + \frac{\Theta_{3,1} }{s \alpha_3} + i0 }
            \Bigg  ]\,,
\\
             & \tilde \Phi = \int  
             \frac{ {\rm d} \alpha_1}{2 \pi i} \frac{\Delta_{4,1}}{
               -s \delta_4 (\alpha_1 + \alpha_4) + \Delta_{4,1}+i 0}
                \Bigg  [\frac{1}{ - \alpha_1 + \frac{\Delta_1}{s}  + i0  }
               + \frac{1}{\alpha_1 + \frac{\Theta_{4,1} }{s \beta_4} + i0 } \Bigg ] \; .
           \end{split} 
          \label{eq:common_int_trick}
           \ee

           It is straightforward to integrate over $\beta_1$ and $\alpha_1$ in Eq.~(\ref{eq:common_int_trick}).  Indeed, focusing on 
           the function $\Phi$, we note that,  if we close  the integration
           contour in the upper half plane, only the residue at $\beta_1 = \Theta_{3,1}/(s\alpha_3)$ contributes. 
           We then find
           \be
\Phi  = (-1) \frac{\Delta_{3,1}}{\Delta_{3,1} +  \delta_3 ( \Theta_{3,1} - s\beta_3) }\,.
           \ee
           Expanding this result  in $ \delta_3$, performing a similar computation
           for $\tilde \Phi$, and keeping only the relevant terms in the product of
           $\Phi$ and $\tilde \Phi$, we obtain  Eq.~\eqref{eq3.30}.

\vspace*{0.5cm}        
Finally, it is convenient to write the one-loop non-factorizable amplitude by extracting 
exact (i.e. not expanded in powers of $\lambda$) Born amplitude. The latter reads 
\be
{\cal M}_0 = ig_W^2\;g_{VVH}
\frac{\langle 3 | \gamma^\mu | 1] \langle 4| \gamma_\mu | 2]}
    {(q_1^2-m_V^2)(q_2^2-m_V^2)}.
    \label{eq.m0}
\ee
Using it, we write 
\be
   {\cal M}_1 = i \frac{g_s^2}{4\pi}\;T^{a}_{i_3 i_1} T^a_{i_4 i_2} \, {\cal M}_0 \, {\cal C}_1,
   \label{eq.amp1}
\ee
 The function $C_1$ reads 
\be
\begin{split} 
{\cal C}_1 = 2 \int &\frac{{\rm d}^{d-2} \bk_{1,\perp} }{(2\pi)^{1-2\epsilon}}
	 \frac{(\bp_{3,\perp}^2+m_V^2)(\bp_{4,\perp}^2+m_V^2)}{\Delta_1 \Delta_{3,1} \Delta_{4,1} } \\
	& \times \left[ 
	1
	-\delta_3 \left(\frac{m_V^2}{\bp_{3,\perp}^2+m_V^2}+\frac{m_V^2}{\Delta_{3,1}}\right)
	-\delta_4 \left(\frac{m_V^2}{\bp_{4,\perp}^2+m_V^2}+\frac{m_V^2}{\Delta_{4,1}}\right)
	\right].
\end{split} 
\label{eq:one_loop_coe_C1}
\ee
We note that the above expression includes both the leading and  the first subleading terms in the
expansion of the one-loop amplitude in powers of $\sqrt{\lambda}$.  
The function ${\cal C}_1$ can be computed analytically and expressed through logarithmic and dilogarithmic functions;
the corresponding discussion can be found in appendix.

\section{Two-loop non-factorizable contributions to WBF}
\label{sec:two_loop_calc}

\begin{figure}[t]
\begin{center}
  \begin{subfigure}{0.48\textwidth}
    \centering
    \includegraphics[height=3.5cm]{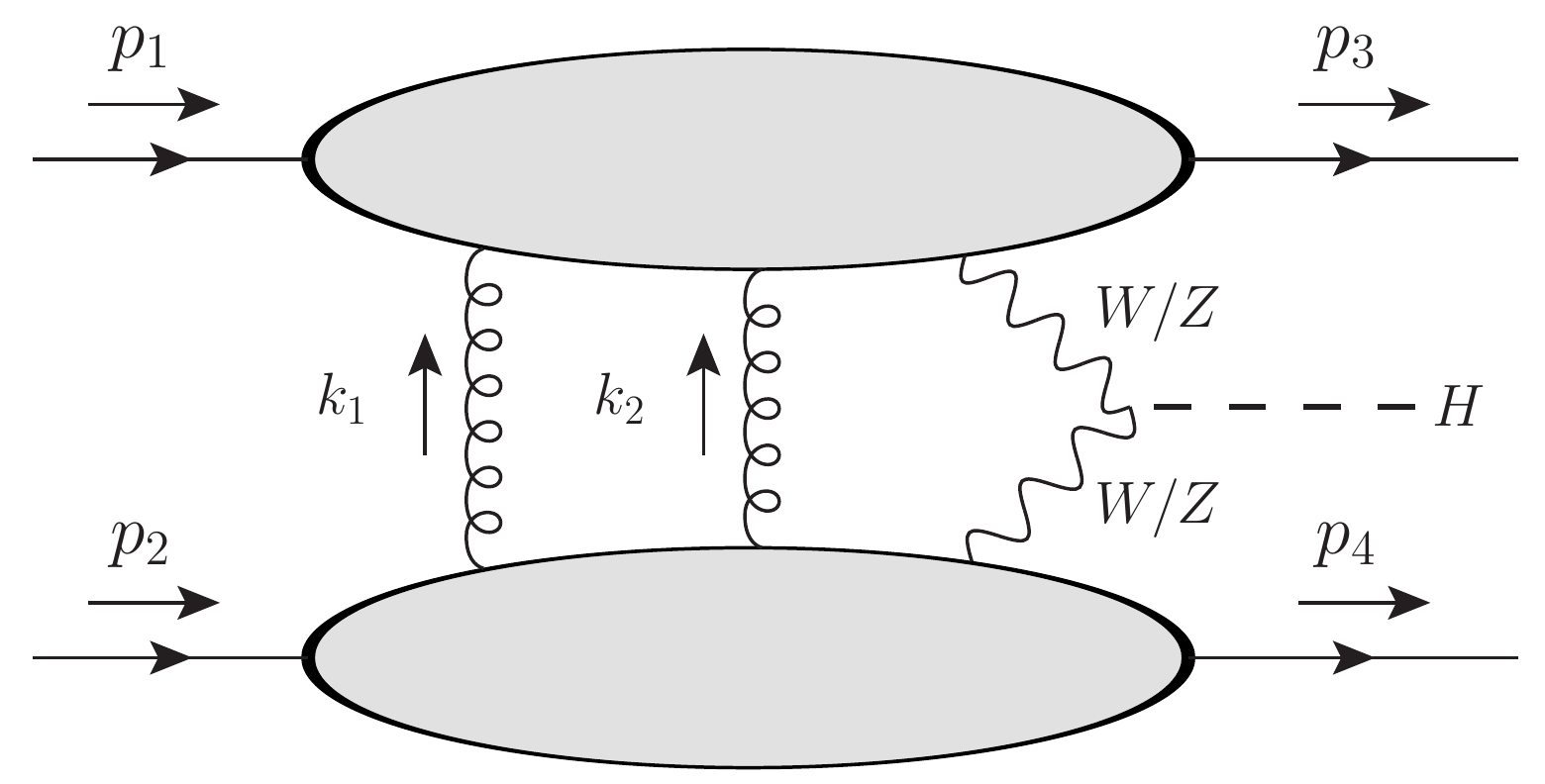}
    \label{fig:wbf_2l}
  \end{subfigure}
  ~
  \begin{subfigure}{0.48\textwidth}
    \centering
    \includegraphics[height=2.8cm]{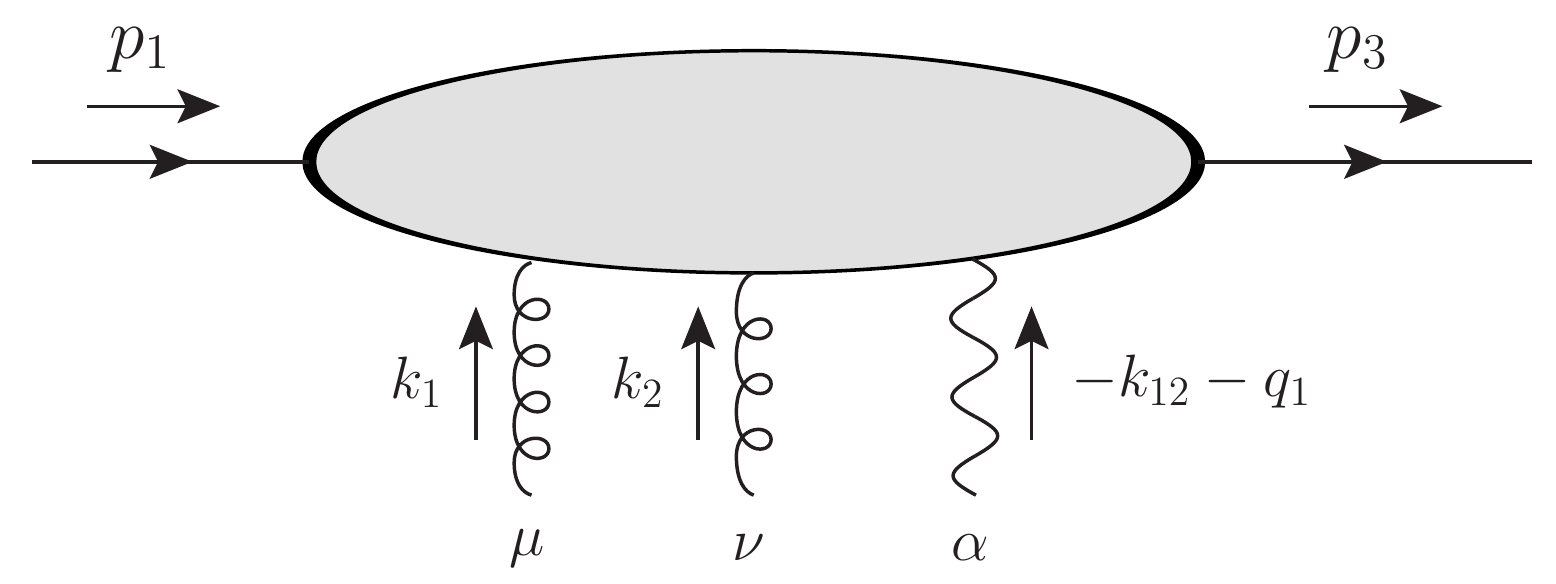}
    \label{fig:current_NNLO}
  \end{subfigure}
  \caption{The two-loop amplitude, shown on the left, is thought in terms of the currents that make it up. On the right, we define of the generalized upper current $J^{\mu \nu \alpha}(k_1,k_2, -k_{12}-q_1)$ used in the calculation of the two-loop amplitude.}
\label{fig1}
\end{center}
\end{figure}

                          We continue with the computation of two-loop non-factorizable
                          QCD corrections to Higgs boson production in weak boson fusion. 
                          The two-loop non-factorizable amplitude is written as 
                          \be
{\cal M}_2 = -i g_s^4 g_W^2 g_{VVH} \left ( \frac{1}{2} \{T^a, T^b \}  \right )_{i_3 i_1}  \left ( \frac{1}{2} \{T^a, T^b \}   \right )_{i_4 i_2} {\cal A}_2\,,  
                          \ee
                          where
                          \be
			  {\cal A}_2 = \frac{1}{2!}\int \frac{{\rm d}^d k_1 }{(2 \pi)^d} \frac{{\rm d}^d k_2 }{(2 \pi)^d}
                             \frac{1}{d_1 d_2 d_3 d_4}
                             J_{\mu \nu \alpha}(k_1,k_2,-k_{12}-q_1) {\tilde J}^{\mu \nu \alpha}(-k_1,-k_2,k_{12} -q_2)\,.
                             \label{eq:two_loop_start}
                          \ee
                          The overall factor $1/2!$ comes from the symmetrization 
			  of two identical gluons and
                          \be
             d_1 = k_1^2 + i0, \;\; d_2 = k_2^2 + i 0,\;\;\; d_3 = (k_{12} +q_1)^2 - m_V^2+ i 0,\;\;\;
                       d_4 = (k_{12} - q_2)^2 - m_V^2 + i 0\,,
                          \ee
are bosonic propagators.\footnote{We use $k_{12} = k_1 + k_2$.}  
Similarly to the one-loop case, in Eq.~\eqref{eq:two_loop_start}, we defined
two quark currents; the conventions are explained  in  \figref{fig1}. The currents read
                          \be
                          \begin{split} 
                            & J^{\mu \nu \alpha}(k_1,k_2,-k_{12}-q_1) = \langle 3 | \Bigg \{
                            \\
&                           \frac{ \gamma^\alpha (\hat p_1 + \hat k_{12}  ) \gamma^\nu (\hat p_1 + \hat k_1) \gamma^\mu}{\rho_1(k_{12})  \rho_1(k_1)}
                          +
                          \frac{ \gamma^\alpha (\hat p_1 + \hat k_{12}  ) \gamma^\mu (\hat p_1 + \hat k_2) \gamma^\nu}{\rho_1(k_{12})  \rho_1(k_2)}
\\
  &                        +
                          \frac{ \gamma^\nu (\hat p_3 - \hat k_{2}  ) \gamma^\alpha (\hat p_1 + \hat k_1) \gamma^\mu}{\rho_3(-k_{2})  \rho_1(k_1)}
                          +
                          \frac{ \gamma^\mu (\hat p_3 - \hat k_1  ) \gamma^\alpha (\hat p_1 + \hat k_2) \gamma^\nu}{\rho_3(-k_1)  \rho_1(k_2)}
\\
   &                       +
\frac{ \gamma^\nu (\hat p_3 - \hat k_{2}  ) \gamma^\mu (\hat p_3 - \hat k_{12} ) \gamma^\alpha}{\rho_3(-k_2)
    \rho_3(-k_{12} )}
                          +
                          \frac{ \gamma^\mu (\hat p_3 - \hat k_{1}  ) \gamma^\nu (\hat p_3 - \hat k_{12}) \gamma^\alpha}{\rho_3(-k_1)  \rho_3(-k_{12})}
                         \Bigg \}  | 1]\,,
                          \end{split} 
                          \label{eq:2l_current}
                          \ee
                          and
                          \be
                          \begin{split} 
                            & {\tilde J}^{\mu \nu \alpha}(-k_1,-k_2,k_{12}-q_2) = \langle 4 | \Bigg \{
                            \\
&                           \frac{ \gamma^\alpha (\hat p_2 - \hat k_{12}  ) \gamma^\nu (\hat p_2 - \hat k_1) \gamma^\mu}{\rho_2(-k_{12})  \rho_2(-k_1)}
                          +
                          \frac{ \gamma^\alpha (\hat p_2 - \hat k_{12}  ) \gamma^\mu (\hat p_2 - \hat k_2) \gamma^\nu}{\rho_2(-k_{12})  \rho_2(-k_2)}
\\
  &                        +
                          \frac{ \gamma^\nu (\hat p_4+ \hat k_{2}  ) \gamma^\alpha (\hat p_2 - \hat k_1) \gamma^\mu}{\rho_4(k_{2})  \rho_2(-k_1)}
                          +
                          \frac{ \gamma^\mu (\hat p_4 + \hat k_1  ) \gamma^\alpha (\hat p_2 - \hat k_2) \gamma^\nu}{\rho_4(k_1)  \rho_2(-k_2)}
\\
   &                       +
\frac{ \gamma^\nu (\hat p_4 + \hat k_{2}  ) \gamma^\mu (\hat p_4 + \hat k_{12} ) \gamma^\alpha}{\rho_4(k_2)
    \rho_4(k_{12} )}
                          +
                          \frac{ \gamma^\mu (\hat p_4 + \hat k_{1}  ) \gamma^\nu (\hat p_4 + \hat k_{12}) \gamma^\alpha}{\rho_4(k_1)  \rho_4(k_{12})}
                         \Bigg \}  | 2]\,.
                          \end{split} 
                          \label{eq:2l_current_tilde}
\ee

To integrate over the loop momenta $k_{1,2}$, for each of them (and also for their linear combinations)
we need to consider  regions  shown in Table~\ref{tab1}. We write
\be
k_i = \alpha_i p_1 + \beta_i p_2 + k_{i,\perp},\;\;\; i=1,2\,.
\ee
The   leading contribution comes from the Glauber region where $\alpha_1 \sim \beta_1 \sim \alpha_2 \sim \beta_2 \sim \lambda$
and  $|\bk_{1,\perp}| \sim | \bk_{2,\perp}| \sim \sqrt{\lambda}$. Similar to the one-loop case,
the leading correction arises 
from the mixed region where some  of the  $\alpha$- or $\beta$-components scale as $\sqrt{\lambda}$. 
Both in the Glauber region and in the mixed region,  the loop momenta
in the numerators of both currents $J^{\mu \nu \alpha}$ and ${\tilde J}^{\mu \nu \alpha}$ can be discarded.
The reason for this is the same  as in the one-loop case and we do not repeat this analysis here.

Building on the experience with the one-loop calculation reported in the
previous section, we can make the following observation. To obtain ${\cal O}(\sqrt{\lambda})$ correction,
we only need to consider the cases where $i)$  one or both $\beta_{1,2}$ components of the loop
momenta  scale as $\sqrt{\lambda}$ and both $\alpha_{1,2}$ scale as $\lambda$,  or $ii)$ the other way around.
If one of the two $\alpha$'s and one of the two $\beta$'s scale as $\sqrt{\lambda}$, then $\alpha_{12}$ and $\beta_{12}$ also
scale as $\sqrt{\lambda}$. As the result, both currents in Eq.~\eqref{eq:2l_current} and Eq.~\eqref{eq:2l_current_tilde}
are suppressed by $\mathcal{O}(\sqrt{\lambda})$. Thus, the
contribution of  this region is suppressed by ${\cal O}(\lambda)$ and can be discarded.  We conclude that, if we want to
construct an integrand which is valid \emph{both} in the Glauber region and in the mixed region, we need to write an
  expression that incorporates $\sqrt{\lambda}$ corrections to one of the currents and that the 
  other current should be taken at leading order.

  These considerations also guide the expansion of the propagators in powers of $\lambda$
  to make them valid in both the Glauber region  and in the mixed region. 
To write the approximate expressions, we define 
                          \be
                          \begin{split} 
                            & \Delta_{i} = - \bk_{i,\perp}^2\,,\; \;\; 
                             \Delta_{3,i} = -(\bk_{i,\perp} - \bp_{3,\perp})^2 - m_V^2\,,
                             \;\;\;\;
                             \Delta_{4,i} = -(\bk_{i,\perp} + \bp_{4,\perp})^2 - m_V^2\,,
                            \\
                            & \Theta_{3,i} = - \left ( \bk_{i,\perp}^2 - 2 \bk_{i,\perp} \cdot \bp_{3,\perp} \right )\,,\;\;\;
  \Theta_{4,i} = - \left ( \bk_{i,\perp}^2 + 2 \bk_{i,\perp} \cdot \bp_{4,\perp} \right )\,, 
                          \end{split}
                          \ee
                          for $i \in \{1,2,12\}$, where $\alpha_{12} = \alpha_1 + \alpha_2$, $\beta_{12} = \beta_1 + \beta_2$ etc. 
and obtain
\begin{align} 
  &
  d_{1,2}
  \approx \Delta_{1,2}+ i0\,,\;\;\; d_3 \approx s \delta_3 (\beta_{12} - \beta_3) + \Delta_{3,12}+ i0\,,
    \;\;\;
\nonumber 
    \\
  & d_4 \approx  -s \delta_4 (\alpha_{12} + \alpha_4) +  \Delta_{4,12}  + i0\,,\;\;  \nonumber
  \\
  &   \rho_1(k_i) \approx s \beta_i + \Delta_i+ i0\,,\;\;\; \rho_3(k_i) \approx s \alpha_3 \beta_3 + \Theta_{3,i} + i 0\,, 
  \\
  &  \rho_2(k_i) \approx s \alpha_i + \Delta_i + i0\,,\;\;\; \rho_4(k_i) \approx s \beta_4 \alpha_i + \Theta_{4,i} + i 0\,. \nonumber
\end{align} 
We emphasize  that the above expressions for
propagators  are valid \emph{both} in   the Glauber region and in the mixed region. Because of that,   we can use
them to compute the two-loop non-factorizable amplitude with ${\cal O}(\sqrt{\lambda})$ accuracy
in the same way as  Eq.~\eqref{eq:common_int_trick} was used to do that in the one-loop case. 

Using the expanded propagators,   we simplify  the currents in Eq.~(\ref{eq:2l_current_tilde})
and write the amplitude as
\be
{\cal A}_2^{a \& b}  =  \frac1{2!}\langle 3 | \gamma^\alpha | 1 ] \langle 4 | \gamma_{\alpha} | 2 ]
   \int \frac{{\rm d}^{d-2} \bk_{1,\perp}}{(2 \pi)^{d-2}} \; \frac{{\rm d}^{d-2} \bk_{2,\perp}}{(2 \pi)^{d-2}}
   \frac{1}{\Delta_1 \Delta_2 \Delta_{3,12} \Delta_{4,12}}  \; \Phi \; \tilde \Phi\,,
   \ee
   where
   \be
\begin{split}
  &   \Phi = \int \frac{{\rm d} \beta_1}{2 \pi i}  \frac{{\rm d} \beta_2}{2 \pi i}
  \frac{\Delta_{3,12}}{s\delta_3(\beta_{12} - \beta_3) + \Delta_{3,12}+i 0}
\Bigg \{
\frac{1 }{( \beta_{12} + \frac{\Delta_{12}}{s} + i0)  (\beta_1 + \frac{\Delta_1}{s} + i 0)}
\\
& +
\frac{1}{( \beta_{12} + \frac{\Delta_{12}}{s} + i0)  (\beta_2 + \frac{\Delta_2}{s}  + i 0)}
                        +
                          \frac{1 }{(-\beta_{2} + \frac{\Theta_{3,2}}{s \alpha_3} +i0)  (\beta_1 + \frac{\Delta_1}{s} + i 0)}
\\
                                         &          +
                          \frac{1 }{(-\beta_{1} + \frac{\Theta_{3,1}}{s \alpha_3} +i0)  (\beta_2 + \frac{\Delta_2}{s} + i 0)}
+\frac{1}{( -\beta_{2} + \frac{\Theta_{3,2}}{s \alpha_3} + i0)  (   -\beta_{12} + \frac{\Theta_{3,12}}{s \alpha_3}  + i 0)}
\\
&   +\frac{1}{( -\beta_{1} + \frac{\Theta_{3,1}}{s \alpha_3} + i0)  (-\beta_{12} + \frac{\Theta_{3,12}}{s \alpha_3}  + i 0)}
\Bigg \}\,,
\label{eq4.9}
\end{split} 
    \ee
and
\be
\begin{split}
  &   \tilde \Phi = \int \frac{{\rm d} \alpha_1}{2 \pi i}  \frac{{\rm d} \alpha_2}{2 \pi i}
  \frac{\Delta_{4,12}}{-s \delta_4 (\alpha_4 + \alpha_{12}) + \Delta_{4,12} + i0
  }
\Bigg \{
\frac{1 }{( -\alpha_{12} + \frac{\Delta_{12}}{s} + i0)  (-\alpha_1 + \frac{\Delta_1}{s} + i 0)}
\\
& +
\frac{1}{( -\alpha_{12} + \frac{\Delta_{12}}{s} + i0)  (-\alpha_2 + \frac{\Delta_2}{s}  + i 0)}
                        +
                          \frac{1 }{(\alpha_{2} + \frac{\Theta_{4,2}}{s \beta_4} +i0)  (-\alpha_1 + \frac{\Delta_1}{s} + i 0)}
\\
                                         &          +
                          \frac{1 }{(\alpha_{1} + \frac{\Theta_{4,1}}{s \beta_4} +i0)  (-\alpha_2 + \frac{\Delta_2}{s} + i 0)}
+\frac{1}{( \alpha_{2} + \frac{\Theta_{4,2}}{s \beta_4} + i0)  (   \alpha_{12} + \frac{\Theta_{4,12}}{s \beta_4}  + i 0)}
\\
&   +\frac{1}{( \alpha_{1} + \frac{\Theta_{4,1}}{s \beta_4} + i0)  (\alpha_{12} + \frac{\Theta_{4,12}}{s \beta_4}  + i 0)}
 \Bigg \}\,.
\end{split}
\label{eq4.11}
    \ee
    To integrate  over $\beta_{1,2}$ and $\alpha_{1,2}$  it is useful to rearrange terms in the curly brackets  in Eqs.~(\ref{eq4.9}, \ref{eq4.11}).
    Focusing on  the  integrand in Eq.~(\ref{eq4.9}), 
    we rewrite it as follows
    \be
    \begin{split}
      & \Bigg \{....  \Bigg \} \to
 \frac{\frac{\Delta_1}{s} + \frac{\Delta_2}{s}-  \frac{\Delta_{12} }{s} }{( \beta_{12} + \frac{\Delta_{12}}{s} + i0)
  (\beta_1 + \frac{\Delta_1}{s} + i 0) (\beta_2 + \frac{\Delta_2}{s} + i 0)}
\\
+ & \frac{\frac{\Theta_{3,1}}{\alpha_3 s} + \frac{\Theta_{3,2}}{\alpha_3 s}-  \frac{\Theta_{3,12} }{\alpha_3 s} }{( -\beta_{12}
      + \frac{\Theta_{3,12}}{\alpha_3 s} + i0)  (-\beta_1 + \frac{\Theta_{3,1}}{\alpha_3 s} + i 0)
       (-\beta_2 + \frac{\Theta_{3,2}}{\alpha_3 s} + i 0) }
\\
+ & 
\left (
\frac{1}{\beta_1 + \frac{\Delta_1}{s} + i 0}
+
\frac{1}{-\beta_{1} + \frac{\Theta_{3,1}}{s \alpha_3} +i0}
\right )\!\!
\left (
\frac{1}{\beta_2 + \frac{\Delta_2}{s} + i 0}
+
\frac{1}{-\beta_{2} + \frac{\Theta_{3,2}}{s \alpha_3} +i0}
\right )\,.
\label{eq4.12}
    \end{split}
    \ee

    We use the above  representation to compute the function $\Phi$ in Eq.~(\ref{eq4.9}). We note that the first
    term in Eq.~(\ref{eq4.12}) can be discarded, because of the location of its poles. Indeed,
    \be
    \begin{split} 
      \Phi_1 & = \int \frac{{\rm d} \beta_1}{2 \pi i}  \frac{{\rm d} \beta_2}{2 \pi i}
        \frac{\Delta_{3,12}}{s\delta_3(\beta_{12} - \beta_3) + \Delta_{3,12}+i 0}
  \\
  & \times
  \frac{\frac{\Delta_1}{s} + \frac{\Delta_2}{s}-  \frac{\Delta_{12} }{s} }{( \beta_{12} + \frac{\Delta_{12}}{s} + i0)
  (\beta_1 + \frac{\Delta_1}{s} + i 0) (\beta_2 + \frac{\Delta_2}{s} + i 0)}
=0\,.  
    \end{split}
    \ee
    To compute the contribution of the second term in Eq.~(\ref{eq4.12}), we close the integration contours in
    the lower half-planes for both integration variables.
    We obtain
    \be
    \begin{split} 
      \Phi_2 & = \int \frac{{\rm d} \beta_1}{2 \pi i}  \frac{{\rm d} \beta_2}{2 \pi i}
        \frac{\Delta_{3,12}}{s\delta_3(\beta_{12} - \beta_3) + \Delta_{3,12}+i 0}
  \\
  & \times
\frac{\frac{\Theta_{3,1}}{\alpha_3 s} + \frac{\Theta_{3,2}}{\alpha_3 s}-  \frac{\Theta_{3,12} }{\alpha_3 s} }{( -\beta_{12}
      + \frac{\Theta_{3,12}}{\alpha_3 s} + i0)  (-\beta_1 + \frac{\Theta_{3,_1}}{\alpha_3 s} + i 0)
      (-\beta_2 + \frac{\Theta_{3,2}}{\alpha_3 s} + i 0) }
\\
& = \frac{ \delta_3 ( \Theta_{3,1}+ \Theta_{3,2} - \Theta_{3,12} )} {\Delta_{3,12}}\,.
    \end{split}
    \ee
    To compute the contribution of the third term in Eq.~(\ref{eq4.12}), we close the integration contours
    for both $\beta_1$ and $\beta_2$ in the upper half-planes. The result reads
    \be
    \begin{split} 
      \Phi_3 & = \int \frac{{\rm d} \beta_1}{2 \pi i}  \frac{{\rm d} \beta_2}{2 \pi i}
        \frac{\Delta_{3,12}}{s\delta_3(\beta_{12} - \beta_3) + \Delta_{3,12}+i 0}
  \\
  & \times
\left (
\frac{1}{\beta_1 + \frac{\Delta_1}{s} + i 0}
+
\frac{1}{-\beta_{1} + \frac{\Theta_{3,1}}{s \alpha_3} +i0}
\right )
\left (
\frac{1}{\beta_2 + \frac{\Delta_2}{s} + i 0}
+
\frac{1}{-\beta_{2} + \frac{\Theta_{3,2}}{s \alpha_3} +i0}
\right )
\\
&=
\frac{\Delta_{3,12}}{s\delta_3(\frac{\Theta_{3,1}}{s} + \frac{\Theta_{3,2}}{s} - \beta_3) + \Delta_{3,12}}
\approx 1 - \frac{\delta_3 ( \Theta_{3,1} + \Theta_{3,2} - s \beta_3 )}{\Delta_{3,12}}\,.
    \end{split}
    \ee
    Adding up $\Phi_{1,2,3}$, we find the following expression for the function $\Phi$ which provides the combined contribution
    of both the Glauber region and the mixed region
    \be
\Phi = \sum \limits_{i=1}^{3} \Phi_{i} = 1 - \frac{\delta_3(\Theta_{3,12} - s\beta_3)}{\Delta_{3,12}}
\,.
    \ee
    The calculation for $\tilde \Phi$ proceeds in an identical way. We obtain 
    \be
\tilde \Phi = 1 - \frac{\delta_3(\Theta_{4,12} - s\alpha_4)}{\Delta_{4,12}}
\,.
    \ee

    Finally, putting everything together and retaining terms that provide ${\cal O}(\sqrt{\lambda})$ corrections,
    we find  the following result for the two-loop non-factorizable amplitude 
    \be
\begin{split} 
    {\cal A}_2^{a \& b} & =  -\frac1{2!}\langle 3 | \gamma^\alpha | 1 ] \langle 4 | \gamma_{\alpha} | 2 ]
   \int \frac{{\rm d}^{d-2} \bk_{1,\perp}}{(2 \pi)^{d-2}} \; \frac{{\rm d}^{d-2} \bk_{2,\perp}}{(2 \pi)^{d-2}}
   \frac{1}{\Delta_1 \Delta_2 \Delta_{3,12} \Delta_{4,12}}
   \\
&    \times 
   \left [ 1 + \frac{ \delta_3 }{\Delta_{3,12}}\left ( s \beta_3 - \Theta_{3,12} \right )  
     +  \frac{ \delta_4 }{\Delta_{4,12}}\left ( s \alpha_4 - \Theta_{4,12} \right )   
     \right ]\,.
   \end{split} 
    \ee

    \vspace*{0.5cm}
    It remains to analyze the contributions of the other regions to the  two-loop non-factorizable amplitude. This analysis proceeds along
    the lines of the discussion of the one-loop case. It  relies on the fact that for soft and collinear
    gluons, fermion  currents simplify dramatically. Consider, for example, the case where $k_1$ is Glauber and $k_2$
    is soft. Naively, this region would contribute at ${\cal O}(\lambda^{-2})$ so that we need to account for subleading
    contributions from this region. In practice, the contribution is ${\cal O}(\lambda)$ suppressed compared to
    a naive estimate.

    Indeed, 
    if $k_2$ is soft and $k_1$ is Glauber, then  $k_{12}$ is also soft.  To understand how the currents
    simplify in this case, consider Eq.~(\ref{eq4.12}). Since $\beta_{12} \sim \beta_{1} \sim \sqrt{\lambda} \gg \lambda$, the
    leading contribution in the last line of  Eq.~(\ref{eq4.12}) vanishes; we then find that 
    the  current in  Eq.~(\ref{eq4.12}) scales as $\lambda^{-1}$, at  variance with the naive scaling $\lambda^{-3/2}$.
    We note that we ignore the pole at $\beta_{1,2}=0$ for the same reason as in the one-loop case, see Eq.~\eqref{eq:ignore_pole}. 
    Since both currents exhibit this behavior, we conclude that the contribution of this
    region to the amplitude scales as ${\cal O}(\lambda^{-1})$ and not as ${\cal O}(\lambda^{-2})$
    as naively expected. For this reason, it  is not relevant
    for the calculation of the two-loop amplitude with the  ${\cal O}(\sqrt{\lambda})$ accuracy.
    
Similar to the one-loop case, we  write  the two-loop amplitude as
\be
   {\cal M}_2 = -\frac{1}{2} \frac{g_s^4}{(4\pi)^2} \left ( \frac{1}{2} \{T^a, T^b \}  \right )_{i_3 i_1}  \left ( \frac{1}{2} \{T^a, T^b \}   \right )_{i_4 i_2} {\cal M}_0 \, {\cal C}_2,
   \label{eq:amp2}
\ee
where  ${\cal M}_0$ is  defined in Eq.~(\ref{eq.m0})   and
the function  ${\cal C}_2$ reads 
\be
\begin{split} 
{\cal C}_2 = 4 \int &
	\frac{{\rm d}^{d-2} \bk_{1,\perp} }{(2\pi)^{1 -2\epsilon}}
	\frac{{\rm d}^{d-2} \bk_{2,\perp} }{\pi(2\pi)^{1-2\epsilon}}
	 \frac{(\bp_{3,\perp}^2+m_V^2)(\bp_{4,\perp}^2+m_V^2)}{\Delta_1 \Delta_2 \Delta_{3,12} \Delta_{4,12} } \\
	& \times \left[ 
	1
	-\delta_3 \left(\frac{m_V^2}{\bp_{3,\perp}^2+m_V^2}+\frac{m_V^2}{\Delta_{3,12}}\right)
	-\delta_4 \left(\frac{m_V^2}{\bp_{4,\perp}^2+m_V^2}+\frac{m_V^2}{\Delta_{4,12}}\right)
	\right].
\end{split} 
\label{eq:two_loop_coe_C2}
\ee
This function looks analogous to the one-loop function  ${\cal C}_1$, c.f. 
Eq.~(\ref{eq:one_loop_coe_C1}).  It is relatively straightforward to
compute ${\cal C}_2$ analytically; the corresponding
discussion can be found in appendix.

\section{Infrared pole cancellation and the finite remainder function} 
\label{sec:pole_cancel}

To compute  the double-virtual non-factorizable contribution to the
differential WBF cross section, we  square the one-loop
amplitude  in Eq.~(\ref{eq.amp1})
and calculate the  interference of the two-loop amplitude in Eq.~(\ref{eq:amp2}) 
  with the Born amplitude.  Summing over spins and colours, we find 
\be
{\rm d}\hat\sigma_{\rm nf}^{\rm NNLO} = \frac{N_c^2-1}{4N_c^2}\,\, \alpha_s^2 \,\, {\cal C}_{\rm nf} \,\, {\rm d}\hat\sigma^{\rm LO},
\ee
where $\alpha_s = g_s^2/4\pi$ is the strong coupling constant,\footnote{Strictly speaking,
  this is the bare coupling constant. However, as we will explain shortly, the function ${\cal C}_{\rm nf}$
  is $\ep$-finite. Because of this, the difference between bare and renormalized coupling
  constants can be ignored.}
${\rm d}\hat\sigma^{\rm LO}$ is the exact  Born differential cross section for Higgs boson production in WBF 
and ${\cal C}_{\rm nf}$ characterizes the non-factorizable corrections. 
The function ${\cal C}_{\rm nf}$ reads 
\be
   {\cal C}_{\rm nf} = {\cal C}_1^2 - {\cal C}_2\,,
\label{eq.cnf}
   \ee
   and all terms that  are suppressed stronger than  ${\cal O}(\sqrt{\lambda})$  are supposed to be discarded
   when computing it.

   We note that functions ${\cal C}_1$ and ${\cal C}_2$  are infra-red divergent; these divergences arise when
   the loop momenta   $\bk_{i,\perp}$, $i=1,2$, vanish. Computing these functions and expanding  in $\epsilon$,
   we find 
\be
\begin{split}
	{\cal C}_1 &= -\frac{1}{\epsilon} + {\cal C}_{1,0} + \epsilon \, {\cal C}_{1,1} + {\cal O}(\epsilon^2)\,,\\
	{\cal C}_2 &= \frac{1}{\epsilon^2} - \frac{2}{\epsilon}\,{\cal C}_{1,0} + {\cal C}_{2,0} + {\cal O}(\epsilon^1)\,.
\end{split}
\ee
Using these results in Eq.~(\ref{eq.cnf}), we obtain 
\be
{\cal C}_{\rm nf} = {\cal C}_{1,0}^2 - 2\,{\cal C}_{1,1} - {\cal C}_{2,0}\,,
\label{eq:two_loop_finite}
\ee
which is infra-red finite and can be computed for $\ep = 0$. The fact that the double-virtual contribution
to non-factorizable corrections in WBF is finite through ${\cal O}(\sqrt{\lambda})$ is in accord with 
Catani's formula for infra-red divergences of generic two-loop amplitudes applied to the WBF process~\cite{Catani:1998bh}. 
Analytic results for the function $C_{\rm nf}$ can be found in the  ancillary file provided with this submission.

\section{Numerical results and phenomenology}
\label{sec:num}

It is instructive to study the results of the calculation in several ways. 
 First, we compare the analytic results
for the function ${\cal C}_{\rm nf}$ at leading order in the $\lambda$-expansion
against numerical results\footnote{We note that very recently
  an analytic result for ${\cal C}_{\rm nf}$ at leading order in the $\lambda$-expansion was computed~\cite{Gates:2023iiv}.}
reported in Ref.~\cite{Liu:2019tuy} and find good agreement. 
Second, to  explore the accuracy of our result in a realistic setting, 
we compare the  one-loop amplitude  including leading and first sub-leading terms in the
$\lambda$-expansion,   with  the \emph{exact} one-loop non-factorizable
amplitude $\mathcal{A}_{1}$. To this end, we generate events that pass the WBF
cuts~\cite{Asteriadis:2021gpd}, use them 
to  evaluate both amplitudes, and  compute the following quantity
\begin{align}
    X_\delta = \frac{\mathcal{A}_{1}-\mathcal{A}_1^{a\& b}}{\mathcal{A}_1^{a\& b}-\mathcal{A}_1^{(0)}}
    \,.
    \label{eq:def_qty}
\end{align}
In Eq.~(\ref{eq:def_qty}), $\mathcal{A}_1$ is the exact amplitude,
$\mathcal{A}_1^{(0)}$ is the leading eikonal amplitude 
\begin{align}
\begin{split} 
{\cal A}_1^{(0)}
 & = 
              - \langle 3 | \gamma^\mu | 1] \langle 4| \gamma_\mu | 2]
                  \;   \int
		  \frac{{\rm d}^{d-2} \bk_{1,\perp} }{(2 \pi)^{d-2}} \; \frac{1}{\Delta_1 \Delta_{3,1} \Delta_{4,1} }
    \,,
\end{split}
\end{align}
and  $\mathcal{A}_1^{a\& b}$ is given in Eq.~(\ref{eq3.30}).  We expect that in WBF kinematics
$X_\delta \sim {\cal O}(\sqrt{\lambda})$ and we would like to check if this is indeed the case.

\vspace*{0.5cm}
WBF events are required to contain at least two jets with transverse momenta $p_{\perp,j} > 25$~GeV
and rapidities $|y_j| < 4.5$. The two jets must have well-separated rapidities, $|y_{j_1} -y_{j_2} | > 4.5$,
and their invariant mass should be larger than $600$~GeV. In addition, the two leading jets must be in
the opposite  hemispheres in the laboratory frame;
this is enforced by requiring that the product of their rapidities in the laboratory frame is negative, $y_{j_1} y_{j_2} < 0$.
Finally, we require that the absolute value of Higgs boson  rapidity  in the \emph{partonic} center-of-mass frame is less than
one,  $|y_H|<1.0$.  We impose this cut  to remove events with too  large
$\delta_3 \sim e^{y_H} $ and $\delta_4 \sim e^{-y_H}$,
see Eq.~\eqref{eq2.6}. We note that the cut on the Higgs rapidity removes just
about $5\%$ of the events that pass  standard   WBF cuts.

\begin{figure}
    \centering
    \includegraphics[width=\textwidth]{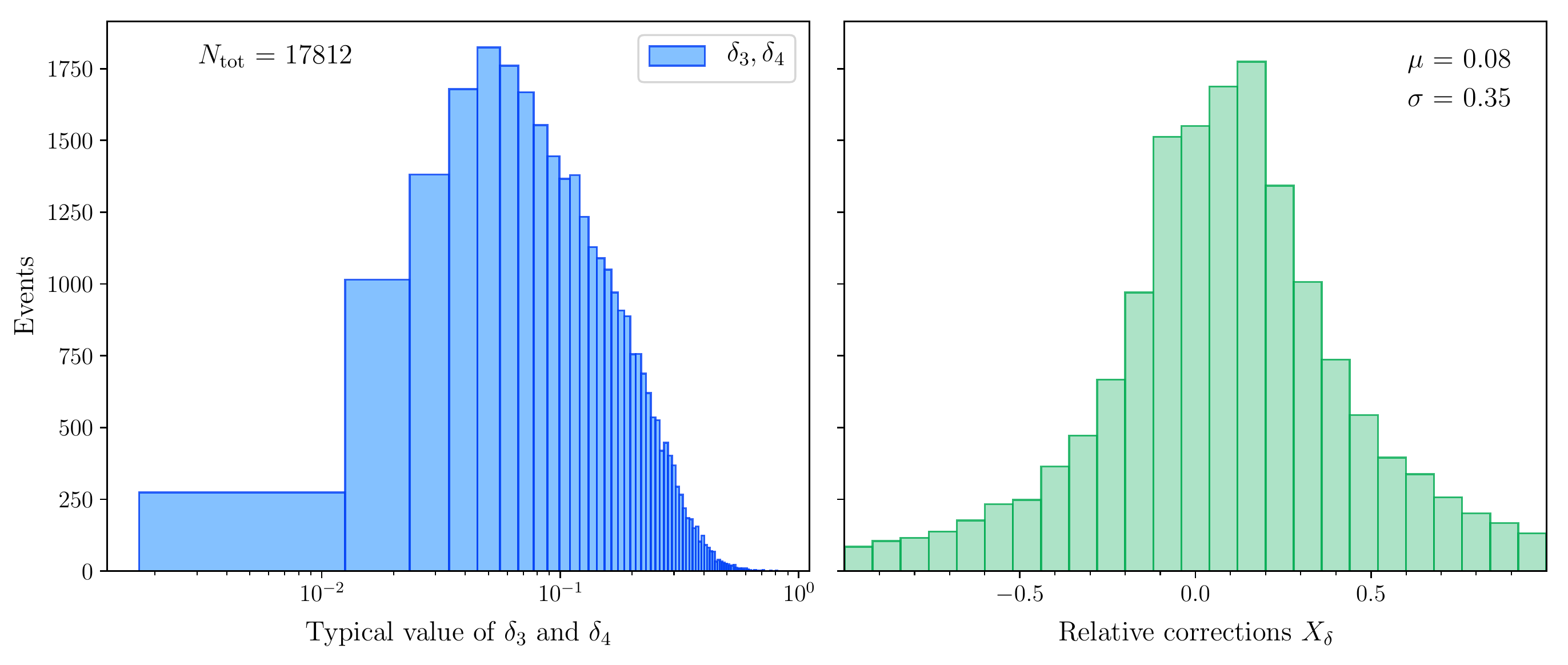}
    \caption{In the left pane,  typical  $\delta_3$ and $\delta_4$ values
      for events allowed by the WBF cuts are shown.
      In the right pane, $X_\delta$ distribution is presented. See text for details. 
}
    \label{fig:check_1l}
\end{figure}

In the left pane in   \figref{fig:check_1l}, we show typical values of $\delta_3$ and $\delta_4$ for selected events.
The distribution peaks at $\delta_3 \sim \delta_4 \sim \sqrt{\lambda} \sim 0.1$ 
which is sufficiently small to justify the expansion in powers
of $\sqrt{\lambda}$. In the right pane in   \figref{fig:check_1l}, we show the distribution of $X_\delta$ 
defined in Eq.~\eqref{eq:def_qty} for selected events.
We see that, on average, the next-to-eikonal corrections reproduce the evaluation of the exact 
one-loop amplitude subject to WBF cuts. The  $X_\delta$-distribution peaks at around $0.1$ which confirms
our expectation that $X_\delta \sim \sqrt{\lambda}$. 
However, the distribution is fairly broad, which means that   neglected terms 
amount  to  about $30\%$ of the \emph{next-to-eikonal contribution}. This is consistent with
magnitude of terms that we neglected by truncating the  $\lambda$-expansion at ${\cal O}(\sqrt{\lambda})$ accuracy.

\begin{figure}[t]
    \centering
      \begin{subfigure}{0.48\textwidth}
    \centering
    \includegraphics[height=5cm]{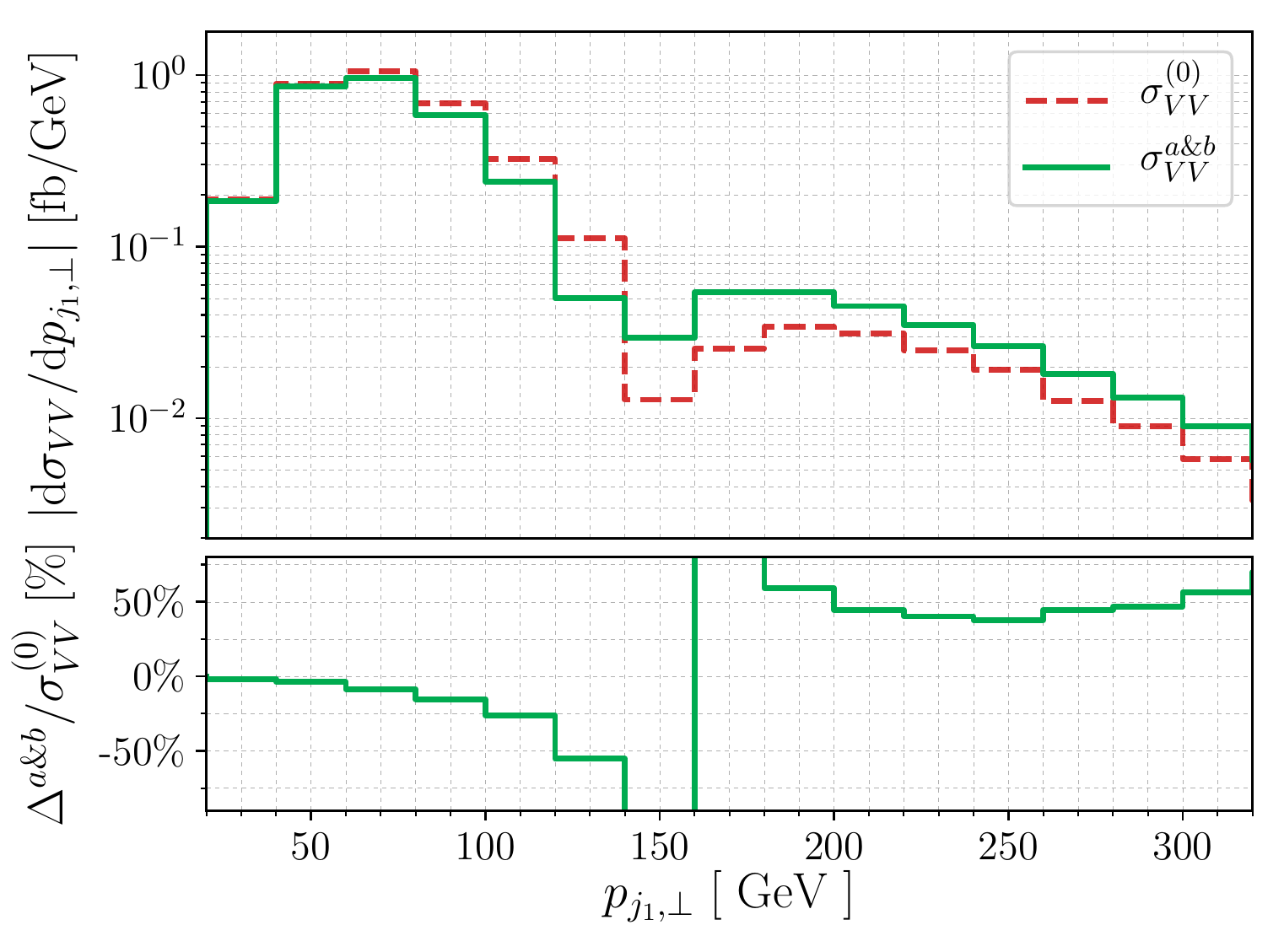}
    \label{fig:ptjet1}
  \end{subfigure}
  ~
  \begin{subfigure}{0.48\textwidth}
    \centering
    \includegraphics[height=5cm]{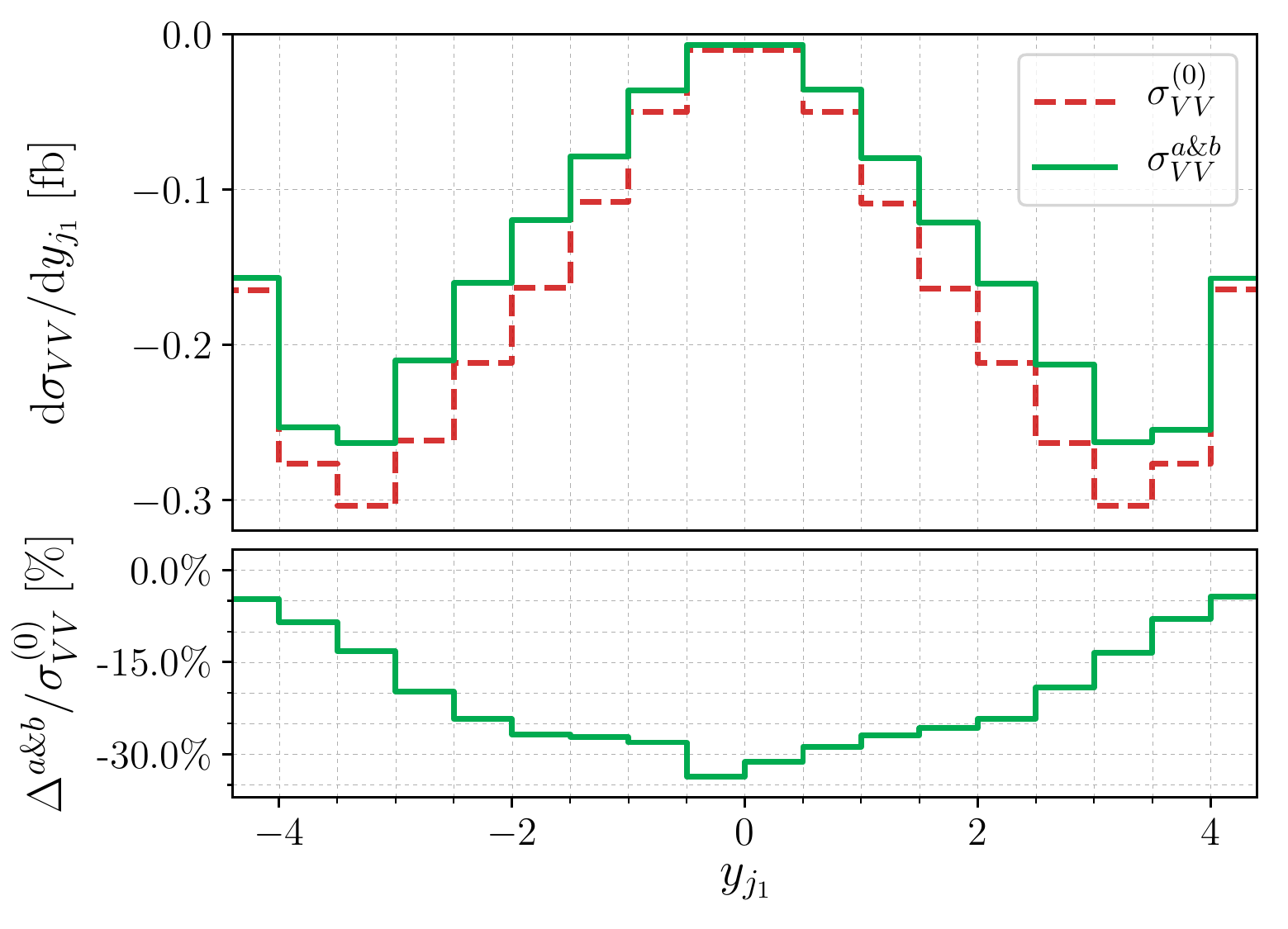}
    \label{fig:yjet1}
  \end{subfigure}
  \caption{Eikonal  and next-to-eikonal contributions to the   transverse momentum  and rapidity distributions of the leading
    jet. In the upper pane,
     leading eikonal contribution is plotted with a red, dashed line
     and the next-to-eikonal one with a green, solid line. In  the lower pane, we show  the ratio of next-to-eikonal to eikonal contributions. 
     We note that in the upper
     left pane, absolute values are shown. See text for further details.}
  \label{fig:distributions_j1}
\end{figure}

\vspace*{0.5cm}
We are now in position to investigate the impact of next-to-eikonal  corrections on the WBF cross section. 
The cross section reads
\begin{align}
    \,d\sigma =  \sum_{i,j}\int \,dx_1 \,dx_2 \, f_i(x_1,\mu_F) \,d\hat{\sigma}_{nf}^{\textrm{NNLO}}(x_1,x_2,\mu_R) \, f_j(x_2,\mu_F)
    \,,
\end{align}
where $f_{i,j}$ are parton distribution functions and $d\hat{\sigma}_{nf}^{\textrm{NNLO}}(x_1,x_2,\mu_R)$ is the partonic
WBF cross section that includes non-factorizable corrections computed through next-to-eikonal approximation. 
We employ  \texttt{NNPDF31\_nnlo\_as\_0118} parton distribution functions~\cite{Buckley:2014ana}
and use dynamical renormalization and factorization scales\footnote{It is not clear that
  this popular choice of the renormalization and factorization scales \cite{Cacciari:2015jma}
  is the optimal choice for non-factorizable contributions.}
\begin{align}
    \mu_F = \mu_R = \frac{m_H}{2} \left[  1 +\frac{ 4 p_{H,\perp}^2}{m_H^2} \right ]^{1/4}
    \,.
\end{align}
We set  the mass of the $W$ boson to  $m_W=80.398 \textrm{ GeV}$, the mass of the $Z$ boson to
$m_Z=91.1876 \textrm{ GeV}$, and the mass of the  Higgs boson to $m_H=125 \textrm{ GeV}$.
The Fermi constant is taken to be $G_F = 1.16637 \times 10^{-5} \textrm{ GeV}^{-2}$.

For $13 \textrm{ TeV}$ proton-proton collisions, we find that the non-factorizable, double-virtual contribution
to  Higgs boson production in WBF evaluates to 
\begin{align}
    \sigma_{VV} = \left(-3.1 + 0.53\right)  \textrm{ fb}
    \,,
\label{eq5.9}
\end{align}
where we display contributions of leading and next-to-leading terms in the $\lambda$-expansion.  We emphasise that  the next-to-eikonal correction is calculated by excluding kinematic configurations where
$|y_H|>1$ in the \emph{partonic} center-of-mass frame, in addition to conventional WBF cuts that we  listed earlier.
It follows from Eq.~(\ref{eq5.9}) that the  correction  to the leading eikonal approximation amounts to  $\mathcal{O}(17\%)$.

\begin{figure}[t]
  \centering
  \begin{subfigure}{0.48\textwidth}
    \centering
    \includegraphics[height=5cm]{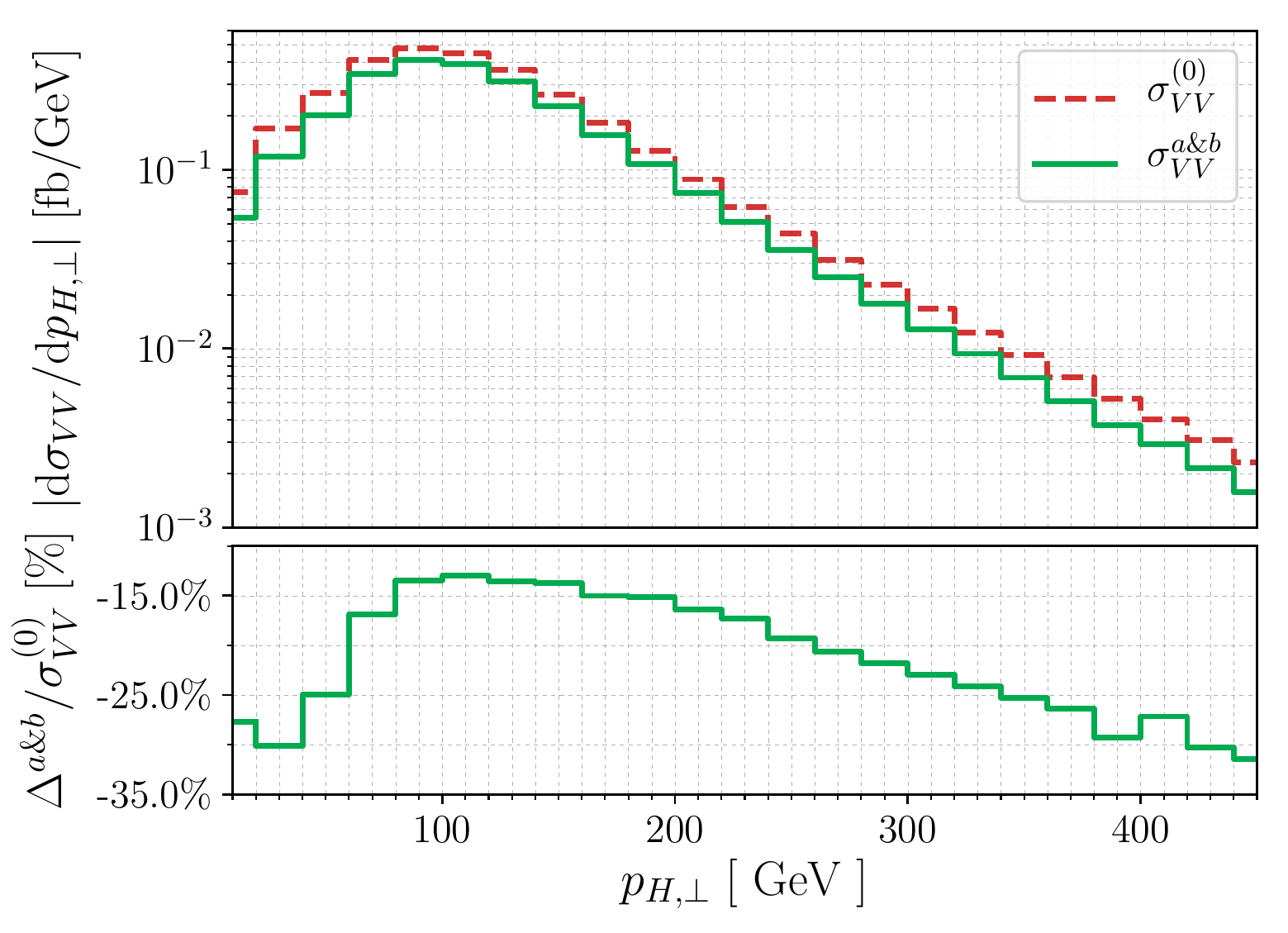}
    \label{fig:ptH}
  \end{subfigure}
  ~
  \begin{subfigure}{0.48\textwidth}
    \centering
    \includegraphics[height=5cm]{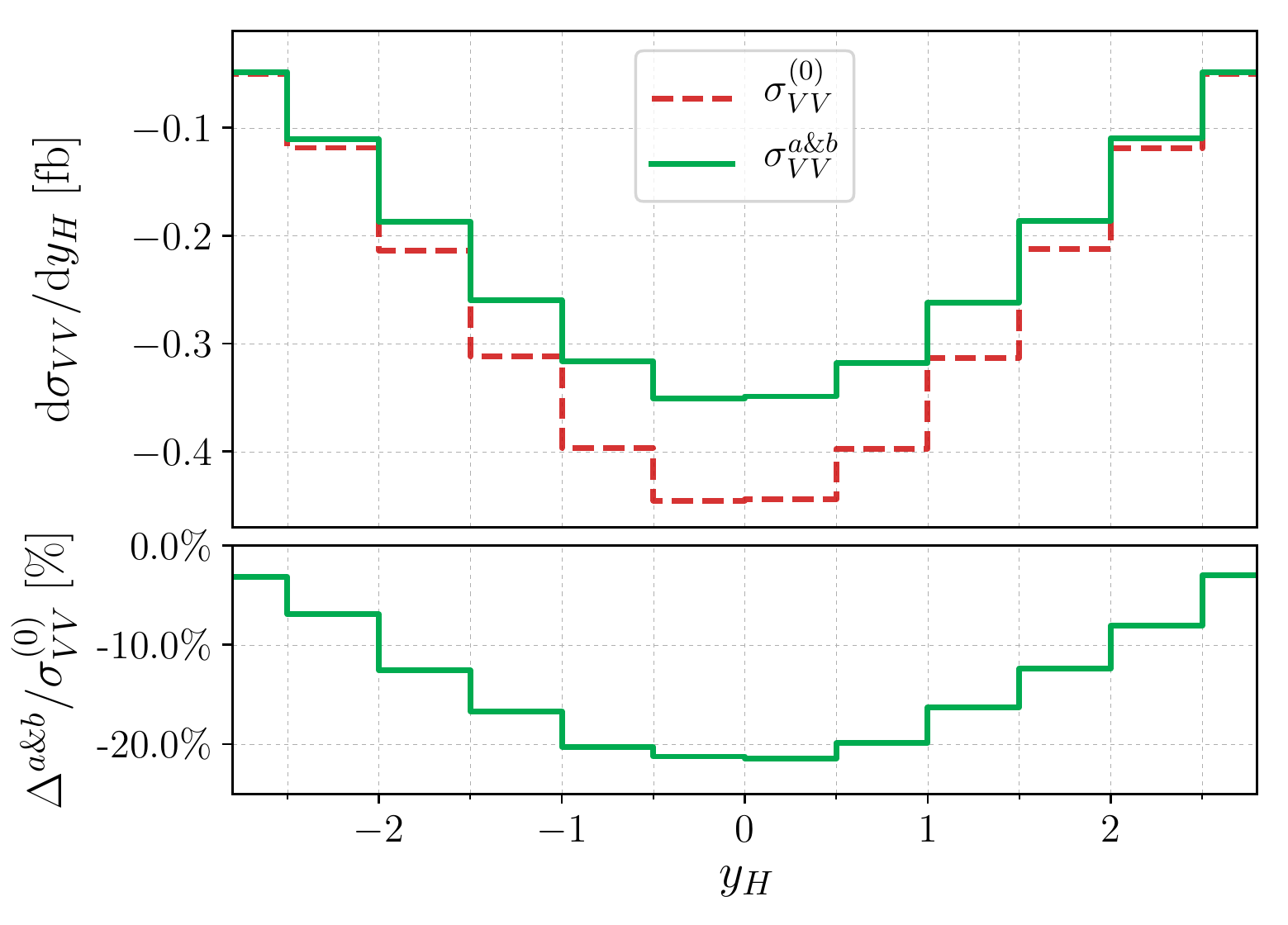}
    \label{fig:yH}
  \end{subfigure}
  \caption{
Eikonal  and next-to-eikonal contributions to the   transverse momentum  and  rapidity distributions of the Higgs 
    boson. In the upper pane,
     leading eikonal contribution is plotted with a red, dashed line
     and the next-to-eikonal one with a green, solid line. In  the lower pane, we show  the ratio of next-to-eikonal to eikonal contributions. 
}
  \label{fig:distributions_H}
\end{figure}

We now turn to the discussion of kinematic distributions. In Fig.~\ref{fig:distributions_j1}, we display non-factorizable
\emph{corrections} to   transverse momentum and rapidity distributions of the leading jet. The comparison of leading
and next-to-leading eikonal contributions in lower panes shows that next-to-leading eikonal corrections
range from ten to fifty percent. They appear to modify the leading order eikonal contribution by ${\cal O}(50\%)$
for higher values of $p_{\perp,j_1}$.  This enhancement is partially related to the fact that the leading eikonal
contribution changes sign at around  $p_{\perp,j_1} \sim 2m_W$, which is the reason for rapidly changing ratio
of eikonal factors shown in the  lower pane. 

The non-factorizable contributions  to Higgs boson transverse momentum and rapidity
distributions are  shown in Fig.~\ref{fig:distributions_H}. The relation between eikonal and next-to-eikonal 
 contributions are similar to what was observed for the fiducial cross section as well as
 $p_\perp$ and rapidity distributions of the leading jet.

\section{Conclusion}
\label{sec:conclusion}

We computed the two-loop virtual non-factorizable QCD corrections to Higgs boson production in weak boson
fusion through   next-to-leading  order in the eikonal expansion.  We found  that such an expansion
proceeds in powers of $p_{\perp, H}/\sqrt{s} \sim m_H/\sqrt{s}$ and explained  how to simplify the integrand of the
two-loop amplitude to calculate  both the  leading   and the next-to-leading terms   in such an expansion.

We observed  that combining individual diagrams before integrating over loop momenta leads to significant simplifications
in  the calculation.   This happens because  contributions of some of the  virtual-momenta regions, 
that are relevant for computing next-to-eikonal corrections in  individual Feynman
diagrams, receive additional suppression in the full amplitude and start contributing only at next-to-next-to-leading
power.

We have derived compact integral representations for the double-virtual
non-factorizable amplitude at both leading and next-to-leading power  in the eikonal expansion. We have
also explained how to compute the two-loop amplitude analytically and provided the analytic results in the ancillary file. 

The numerical impact of next-to-eikonal corrections is significant although, given the overal smallness of non-factorizable
contributions, they do not change the original conclusions of Refs.~\cite{Liu:2019tuy,Dreyer:2020urf}. Nevertheless, we find that, typically, the next-to-eikonal
corrections  change the estimate of the non-factorizable contributions based on the leading term in the eikonal expansion by
${\cal O}(20)$ percent.  

As a final comment, we note that other  sources of non-factorizable contributions to WBF cross sections,
including double-real emission and the real-virtual corrections,   were recently studied in Ref.~\cite{Asteriadis:2023nyl}.
It was found that, thanks to the WBF cuts,   all the contributions beyond the double-virtual ones are tiny and
cannot impact the phenomenological studies  of Higgs production in WBF in any way.  The results reported in
this reference allow us to estimate the contribution of the  non-factorizable double-virtual corrections to the WBF
cross section with  a precision that is likely better than ${\cal O}(10)$ percent. Since the non-factorizable contribution itself is just ${\cal O}(1)$ percent of the total WBF cross section, the remaining uncertainties stemming from the
imprecise knowledge of the two-loop virtual amplitude are irrelevant. We conclude that the current
understanding of non-factorizable effects  is sufficient for phenomenological studies of Higgs production
in weak boson fusion envisaged for the 
Run III and the high-luminosity phase of the LHC.

\section{Acknowledgments}
We would like to thank A. Penin for useful conversations about non-factorizable effects in Higgs production in WBF. 
We  are grateful to K.~Asteriadis and Ch.~Br\o{}nnum-Hansen for their help
with the implementation of next-to-eikonal corrections into a numerical code
for computing non-factorizable contributions to the WBF cross section. This research is partially supported
by the Deutsche Forschungsgemeinschaft (DFG, German Research Foundation) under
the grant 396021762 - TRR 257. The diagrams in Figs.~\ref{fig0} and \ref{fig1} were generated using \texttt{Jaxodraw}~\cite{Binosi:2003yf}.

\appendix
\section{Calculation of two-dimensional master integrals}

The goal of this appendix is to   explain how  the $d=2$  Feynman integrals 
that contribute to the coefficients $\mathcal{C}_{1,2}$ can be computed. 
We begin with the discussion  of the two-loop case. Two-loop 
$d=2$ integrals that are required for computing $\mathcal{C}_2$ 
belong to  the following integral family
\be
j[a_1, a_2, a_3, a_4] = \frac{(m_V^2)^{2\epsilon}}{\pi^{d-2}\Gamma(1+\epsilon)^2}\int 
\frac{{\rm d}\bk_{1,\perp}^{d-2}{\rm d}\bk_{2,\perp}^{d-2}}{\Delta_1^{a_1}\Delta_2^{a_2}\Delta_{3,12}^{a_3}\Delta_{4,12}^{a_4}}.
\ee
These integrals depend  on the transverse momenta of the outgoing jets and of the Higgs boson, as well
as on the mass of the vector boson $V$. For later convenience, we introduce three dimensionless variables as
\be
x = \frac{\bp_{3,\perp}^2}{m_V^2}, \quad
y = \frac{\bp_{4,\perp}^2}{m_V^2}, \quad
z = \frac{\bp_{H,\perp}^2}{m_V^2}. \quad
\ee

\vspace*{0.4cm}
It is straighforward to write down integration-by-parts (IBP) identities \cite{Tkachov:1981wb,Chetyrkin:1981qh} for the integral
family $j[a_1,a_2,a_4,a_4]$.
Performing the IBP reduction with {\sc LiteRed} \cite{Lee:2012cn,Lee:2013mka},  we find  that there
are six  master integrals. They are 
\be
\begin{aligned}
	f_1 &= j[2,1,2,0], & f_2 &= j[2,2,1,0], & f_3 &= j[2,1,0,2], \\
	f_4 &= j[2,2,0,1], & f_5 &= j[2,1,1,1], & f_6 &= j[2,1,2,1].
\end{aligned}
\ee

\begin{figure}[t]
  \centering
  \begin{subfigure}{0.2\textwidth}
    \centering
    \includegraphics[width=\textwidth]{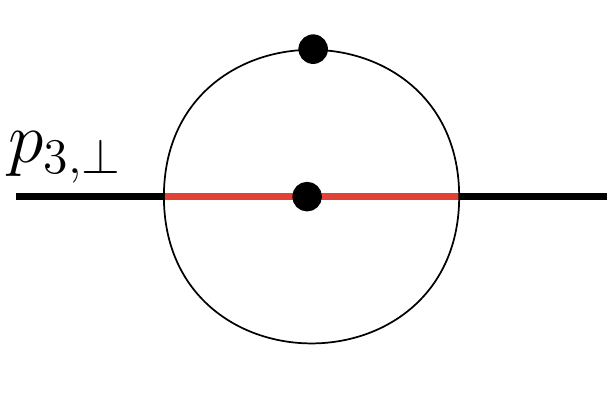} 
  \caption*{$f_1$}
  \end{subfigure}
  ~
  \begin{subfigure}{0.2\textwidth}
    \centering
    \includegraphics[width=\textwidth]{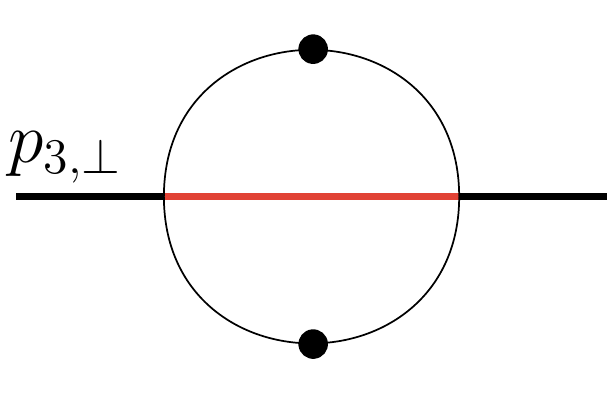} 
  \caption*{$f_2$}
  \end{subfigure}
  ~
  \begin{subfigure}{0.2\textwidth}
    \centering
    \includegraphics[width=\textwidth]{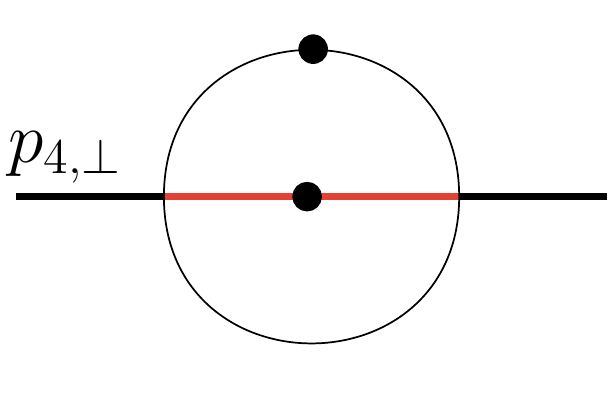} 
  \caption*{$f_3$}
  \end{subfigure}
  ~
  \begin{subfigure}{0.2\textwidth}
    \centering
    \includegraphics[width=\textwidth]{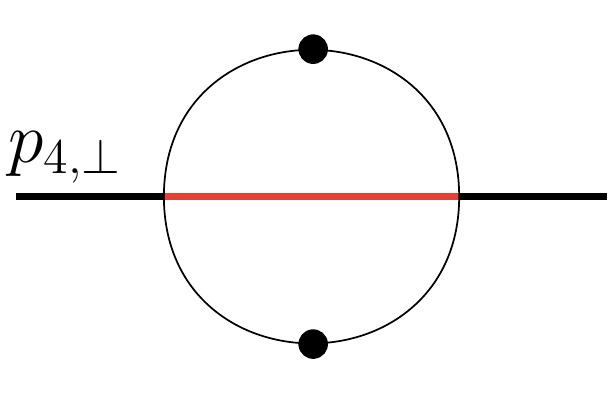} 
  \caption*{$f_4$}
  \end{subfigure}
	\\
  \begin{subfigure}{0.2\textwidth}
    \centering
    \includegraphics[width=\textwidth]{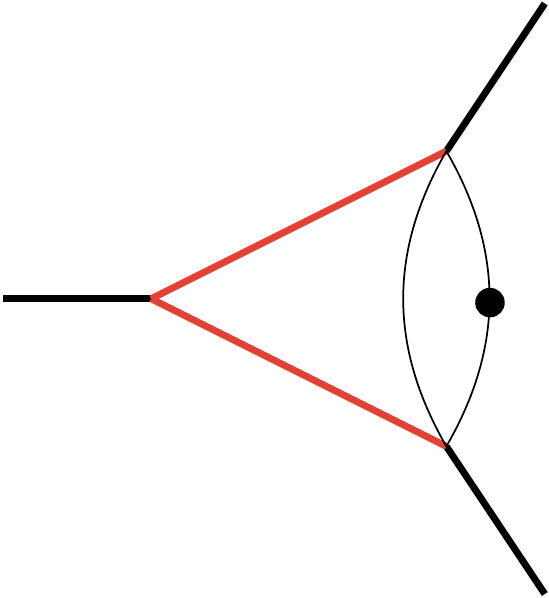} 
  \caption*{$f_5$}
  \end{subfigure}
  \qquad\qquad
  \begin{subfigure}{0.2\textwidth}
    \centering
    \includegraphics[width=\textwidth]{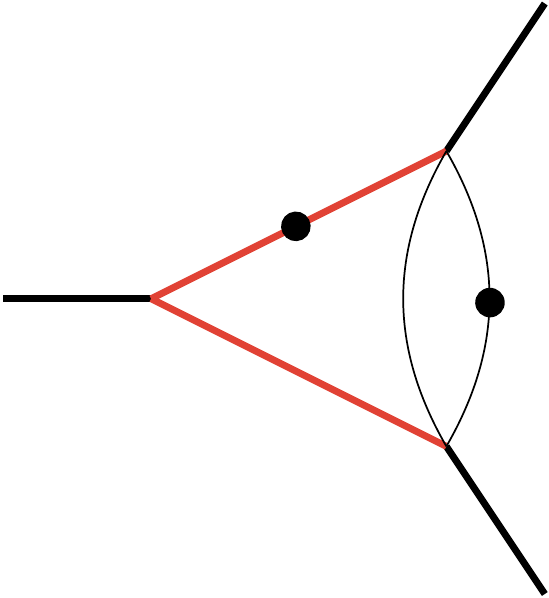} 
  \caption*{$f_6$}
  \end{subfigure}
\caption{Two-dimensional two-loop master integrals. The thick and thin internal lines 
	represent massive and massless propagators, respectively. Red lines have
	mass $m_V$. The black thick lines correspond to external "massive" legs.
	A dot on the internal line means raising the power of corresponding propagator by one.}
\label{fig:2dim_mis}
\end{figure}

The master integrals  are displayed in  Fig. \ref{fig:2dim_mis}. Although we need these integrals at $d=2$,
we find it more convenient to study them first in  four dimensions. In particular,
at $d=4$,  we easily obtain the  canonical basis~\cite{Henn:2013pwa}  using the Magnus series expansion method \cite{Argeri:2014qva}.
We then  transform  the integrals to $d=2$ using the  
dimensional recurrence relations \cite{Tarasov:1996br}. 
In four dimensions, the canonical basis reads 
\begin{align}
	g_1 & = x \epsilon ^2 m_V^2 f_1 , \nn
	g_2 & = 2 \epsilon ^2 m_V^2 f_1+(x+1) \epsilon ^2 m_V^2 f_2, \nn
	g_3 & = y \epsilon ^2 m_V^2 f_3 , \nn
	g_4 & = 2 \epsilon ^2 m_V^2 f_3 +(y+1) \epsilon ^2 m_V^2 f_4, \nn
	g_5 & = 2 \epsilon ^3 m_V^2 r_2 f_5 , \\
	g_6 & = \frac{\epsilon ^2 r_1 m_V^2}{4 [2(y-x)+z(1+y)]}
	\Bigg\{
	4 m_V^2[(x-y)^2-(x+1) (y+1) z] f_6 \nn
	&\qquad -6 \epsilon [(x-y)(y-1)+z(1+y)]f_5
	+(y+1)^2 (f_4+2f_3) \nn
	&\qquad -(x+1) (y+1) (f_2+2f_1) 
	\Bigg\}, \nonumber
\end{align}
where $r_{1,2}$ represent two square roots,
\be
r_1 = \sqrt{z(z+4)}, \qquad
r_2 = \sqrt{(x+y-z)^2-4xy}.
\ee
Note that all the $g$'s are normalized to be dimensionless and can be regarded as  
functions of $x,y$ and $z$ only. The canonical basis vector $\vec{g} =(g_1,g_2,g_3,g_4,g_5,g_6)^T$ satisfies a differential
equation in the $d{\rm log}$ form,
\be
d \vec{g}(x,y,z;\epsilon) = \epsilon (d\mathbb{A}) \; \vec{g}(x,y,z;\epsilon),
\label{eq.app1}
\ee
where the matrix $\mathbb{A}$ reads
\be
\scalemath{0.9}{
\left(
\begin{array}{cccccc}
 l_1-2 l_2 & -l_2 & 0 & 0 & 0 & 0 \\
 4 \left(l_1-l_2\right) & -2 l_2 & 0 & 0 & 0 & 0 \\
 0 & 0 & l_3-2 l_4 & -l_4 & 0 & 0 \\
 0 & 0 & 4 \left(l_3-l_4\right) & -2 l_4 & 0 & 0 \\
 2 l_{10}-l_{12}+l_{14} & \frac{l_{14}-l_{12}}{2}  & 2 l_{11}+l_{12}-l_{14} & \frac{l_{12}-l_{14}}{2}  & \frac{l_5-3 l_7+4 l_9}{2} & -2 \left(l_{12}+l_{14}\right) \\
 \frac{-l_8+l_{12}-2 l_{13}+l_{14}}{4}  & \frac{2 l_8+l_{12}-2 l_{13}+l_{14}}{8}  & \frac{-l_8-l_{12}+2 l_{13}-l_{14}}{4}  & \frac{2 l_8-l_{12}+2 l_{13}-l_{14}}{8}  & -\frac{l_{12}+l_{14}}{8}  & \frac{l_5-2 l_6-l_7}{2}  \\
\end{array}
\right),
}
\ee
and the 14 logarithms that constitute $\mathbb{A}$ are
\begin{align}
\label{eq:app_letters}
	l_1 &= \log (x), &
	l_2 &= \log (x+1), &
	l_3 &= \log (y), \nn
	l_4 &= \log (y+1), &
	l_5 &= \log (z), &
l_6 &= \log (z+4), \nn
	l_7 &= \rlap{$\displaystyle
	\log \left[(x-y)^2-(x+1) (y+1) z\right], \,
l_8 = \log \left(\frac{z+2-r_1}{z+2+r_1}\right), \,
l_9 = \log \left(r_2\right),
	$} \nn
	l_{10} &= \log \left(\frac{x-y+z-r_2}{x-y+z+r_2}\right), &
	l_{11} &= \log \left(\frac{-x+y+z-r_2}{-x+y+z+r_2}\right), \\
	l_{12} &= \log \left(\frac{r_1-r_2+x-y}{r_1+r_2+x-y}\right), &
l_{13} &= \log \left(\frac{r_1-r_2+x-y}{r_1+r_2-x+y}\right), \nn
l_{14} &= \log \left(\frac{r_1-r_2-x+y}{r_1+r_2-x+y}\right). \nonumber
\end{align}
Eq. (\ref{eq.app1}) can be recursively solved order-by-order in $\epsilon$ and
the solutions are expressed in terms
of Chen's iterated integrals \cite{Chen:1977oja} with some boundary constants that cannot
be determined from the differential equations alone.   For the integrals $g_{1,..,6}$
these constants can be computed with a relative ease since all canonical integrals,
except  $g_2$ and $g_4$,  vanish when $x = y = z = 0$.  The non-vanishing
integrals $g_2$ and $g_4$ at this kinematic point evaluate to 
\be
g_{2,4}(0,0,0) = -\frac{\pi  \epsilon  \csc (\pi  \epsilon ) \Gamma (1+2 \epsilon)}{\Gamma (1+\epsilon)^2}
=-1- \frac{\pi^2}{3}\epsilon ^2+\mathcal{O}\left(\epsilon ^3\right).
\ee
Furthermore, under the change of variables
\be
x=zuv, \quad y=z(1-u)(1-v), \quad z=\frac{(1-w)^2}{w},
\ee
the square roots $r_{1,2}$ are rationalized simultaneously and we find 
\be
r_1 = \frac{(1-w)(1+w)}{w}, \qquad
r_2 = \frac{(1-w)^2(u-v)}{w}.
\ee
As the result,  the solutions of the system Eq.~(\ref{eq.app1})
can be expressed in terms of  multiple polylogarithms.
In fact,  since we need $\vec{g}$ only through  ${\cal O}(\epsilon^2)$, relevant expressions for
integrals involve logarithms and dilogarithms of $u,v,w$. To express them in terms of $x,y,z$,
we use the following formulas 
\be
u,v = \frac{x-y+z\pm r_2}{2z}, \quad w = \frac{2+z-r_1}{2}.
\ee

Finally, to compute the one-loop amplitude, we need to study the following integral family
\be
j[a_1, a_2, a_3] = \frac{(m_V^2)^{\epsilon}}{\pi^{(d-2)/2}\Gamma(1+\epsilon)}\int 
\frac{{\rm d}\bk_{1,\perp}^{d-2}}{\Delta_1^{a_1}\Delta_{3,1}^{a_2}\Delta_{4,1}^{a_3}}.
\ee
The analysis is identical to the two-loop case and we will not repeat it here. 
We only mention that  the canonical basis at $d=2$ reads 
\begin{align}
g_1 &= \epsilon j[0,0,1], \nn
g_2 &= \epsilon j[0,1,1]m_V^2 r_1 , \nn
g_3 &= \epsilon j[1,0,1]m_V^2\left(1+y\right), \nn
g_4 &= \epsilon j[1,1,0]m_V^2\left(1+x\right), \nn
g_5 &= \frac{\epsilon m_V^2}{2 r_2}
	\Big\{
	2 j[1,1,1]m_V^2\left[(x-y)^2-(1+x)(1+y)z\right] \\
	&\qquad +j[1,1,0]\left[\left(-x+y-z\right)+x\left(x-y-z\right)\right] \nn
	&\qquad -j[1,0,1]\left[\left(-x+y+z\right)+y\left(x-y+z\right)\right] \nn
	&\qquad +j[0,1,1]\left(2-x-y+z\right)z
	\Big\}. \nonumber
\end{align}
The canonical basis statisfies a differential equation in the $d{\rm log}$ form, similar to Eq.~\eqref{eq.app1}. The corresponding matrix $\mathbb{A}$ reads
\be
\scalemath{0.9}{
\left(
\begin{array}{ccccc}
 0 & 0 & 0 & 0 & 0 \\
 l_8 & -l_6 & 0 & 0 & 0 \\
 -l_3 & 0 & l_3-2 l_4 & 0 & 0 \\
 -l_1 & 0 & 0 & l_1-2 l_2 & 0 \\
 \frac{l_{10}+l_{11}}{2}  & \frac{l_{12}+l_{14}}{2} & \frac{-l_{11}-l_{12}+l_{14}}{2}
    & \frac{-l_{10}+l_{12}-l_{14}}{2}  & 2 l_9-l_7 \\
\end{array}
\right),
}
\ee
where the logarithms, $l_i$, are given in Eq. (\ref{eq:app_letters}).
To compute the boundary constants, we use the fact that the basis is finite at $x=y=z=0$ and $g_1 = -1$.

\bibliographystyle{JHEP}
\bibliography{references}

\end{document}